\documentclass[aps,prb,superscriptaddress,amsmath,amssymb,twocolumn,a4paper,floatfix]{revtex4-1}
\usepackage{graphicx}
\usepackage{float}
\usepackage{hyperref}
\usepackage{bm}
\usepackage{tabularx}
\usepackage{xcolor}
\usepackage{bbm}
\usepackage{accents}

\hypersetup{
	breaklinks=true,   
	colorlinks=true,   
}

\DeclareMathOperator{\tr}{tr}

\begin{document}
\title{Pseudoparticle vertex solver for quantum impurity models}
\author{Aaram J. Kim}
\affiliation{Department of Physics, University of Fribourg, 1700 Fribourg Switzerland}
\author{Jiajun Li}
\affiliation{Department of Physics, University of Fribourg, 1700 Fribourg Switzerland}
\affiliation{Paul Scherrer Institute, Condensed Matter Theory, 5352 PSI Villigen, Switzerland}
\author{Martin Eckstein}
\affiliation{Department of Physics, University of Erlangen-N\"urnberg, 91058 Erlangen, Germany}
\author{Philipp Werner}
\affiliation{Department of Physics, University of Fribourg, 1700 Fribourg Switzerland}

\begin{abstract}
	We present a quantum impurity solver based on a pseudo-particle framework, which combines diagrammatic resummations for a three-point vertex with diagrammatic Monte Carlo sampling of a four-point vertex. This recently proposed approach [A. J. Kim et al., arXiv:2112.15549] is generalized here to fermionic impurity problems and we discuss the technical details of the implementation, including the time-stepping approach, the Monte Carlo updates, and the routines for checking the two-particle irreducibility of the four-point vertex. We also explain how the vertex information can be efficiently stored using a Dubiner basis representation. The convergence properties of the algorithm are demonstrated with applications to exactly solvable impurity models and dynamical mean field theory simulations of the single-orbital Hubbard model. 
It is furthermore shown that the algorithm can handle a two-orbital problem with off-diagonal hybridizations, which would cause a severe sign problem in standard hybridization-expansion Monte Carlo simulations. 
Since the vertex-based algorithm successfully handles sign oscillating integrals in equilibrium and samples only connected diagrams, it may be a promising approach for real-time simulations.
\end{abstract}

\maketitle

\section{Introduction}
Quantum impurity models play an important role as representations of correlated atoms in metallic hosts or quantum dots, and as auxiliary systems 
within dynamical mean field theory (DMFT)~\cite{Georges1996} for interacting lattice models.
The models consist of an interacting impurity subsystem with a finite-dimensional Hilbert space embedded in a large uncorrelated particle bath, which generically prevents an exact solution.
A broad range of numerical techniques has been developed to treat quantum impurity problems, 
including exact diagonalization (ED),\cite{Caffarel1994} tensor network approaches,\cite{Wolf2014,Bauernfeind2017} numerical renormalization group (NRG),\cite{Bulla2008} and quantum Monte Carlo (QMC)\cite{Hirsch1986,Gull2011} methods. 
Because of the relatively low computational cost, perturbative expansions have also been widely adopted in model calculations, including
the iterated perturbation theory,\cite{Kajueter1996} slave-particle schemes\cite{Coleman1984,Kotliar1986,Li1989} and self-consistent strong-coupling expansions.\cite{Keiter1971,Pruschke1989} For many equilibrium problems, state-of-the-art QMC methods such as the continuous-time interaction expansion (CT-INT)~\cite{Rubtsov2005} and continuous-time hybridization expansion (CT-HYB)~\cite{Werner2006} enable efficient simulations of relevant multi-orbital or cluster-impurity problems at not too low temperatures. Recent progress in the development of NRG-based solvers allows to 
solve these models 
at arbitrarily low temperatures.\cite{Toth2008,Weichselbaum2012,Mitchell2014} 

The development of quantum impurity solvers for nonequilibrium applications, such as quantum dots with an applied voltage bias, or the impurity problems which need to be solved within the nonequilibrium extension of DMFT,\cite{Aoki2014} remains an active and challenging research frontier. Adaptations of the NRG,\cite{Anders2006,Joura2008}
MPS,\cite{Wolf2014b,Balzer2015}
and QMC,\cite{Muehlbacher2008,Werner2009,Werner2010} methods have enabled the study of specific problems, but the existing implementations of numerically exact methods are restricted to models with small local Hilbert spaces or to short-time simulations. In particular, methods like CT-INT and CT-HYB suffer from a dynamical sign problem, which grows exponentially with the length of the simulated real-time interval.\cite{Werner2009} Hence, 
both for nonequilibrium DMFT applications and for the calculation of real-frequency spectra 
approximate impurity solvers such as the non-crossing approximation\cite{Keiter1971} (NCA) have been frequently employed. NCA is based on a pseudo-particle formalism\cite{Barnes1976,Coleman1984} and may be regarded as the first-order approximation of the pseudo-particle self-energy in the hybridization function. If this self-energy is inserted into the pseudo-particle Dyson equation, it generates the subset of CT-HYB diagrams without crossing hybridization lines. 
NCA preserves the energy and particle conservation laws, but the results are qualitatively correct only in the strongly correlated regime.
By extending the approach to higher-order expansions, e.g., 
the one-crossing approximation \cite{Pruschke1989} and third-order schemes,\cite{Eckstein2010} one can systematically improve its numerical 
accuracy. However, 
the computational cost significantly increases with increasing expansion order, again limiting the applicability to small clusters and short simulation times.

To overcome the above-mentioned difficulties, 
two different routes have been explored to improve the accuracy of pseudo-particle based impurity solvers. One strategy is to introduce a three-point vertex in the diagrammatic expression of the pseudo-particle self-energy. The self-consistent calculation of this vertex in the ``symmetrized finite-$U$ NCA" (SUNCA) approach~\cite{Haule2001} allows to sum up additional classes of hybridization-expansion diagrams, for example vertex corrections with an alternating sequence of local states.
The second route which has been successfully explored is to combine the time-stepping scheme for the pseudo-particle propagators~\cite{Eckstein2010} with a Monte Carlo sampling of the pseudo-particle self-energies.  
This so-called inchworm algorithm~\cite{Cohen2015} overcomes the serious dynamical sign problem of CT-INT and CT-HYB by sampling the self-energy instead of the partition function, and in principle enables numerically exact solutions of real-time quantum impurity problems. 
The high numerical cost, however, has so far limited the application of this approach within DMFT to relatively short times.\cite{Dong2017}

In the present work, we combine key aspects of the latter two pseudo-particle methods by developing an impurity solver which supplements a 
diagrammatic calculation of a three-point vertex with a diagrammatic Monte Carlo (diagMC)~\cite{Prokofev1998,VanHoucke:2010ky,Kozik:2010fla} sampling of a 
four-point vertex.\cite{Aaram2021}
 Upon convergence of this self-consistent scheme, and of the diagMC simulation in powers of the hybridization function, 
 a numerically exact solution of the impurity model is obtained. 
We discuss the technical details of the implementation of this vertex-based solver and demonstrate its  
properties with applications to the equilibrium Anderson impurity model (AIM). We find that the diagMC sampling of the four-point vertex yields accurate results despite the oscillating signs in the integrals. In particular, the vertex-based algorithm remains numerically stable in the presence of off-diagonal hybridization functions, which can create a severe sign problem in CT-HYB.\cite{Eidelstein2020} This makes the method promising for
equilibrium applications with sign problems, such as DMFT solutions of spin-orbit coupled systems.\cite{Kim2017,Kim2020a} 
Moreover, because the diagrams for the four-point vertex are connected, the algorithm is also likely to tackle the dynamical sign problem on the real-time axis, in a way similar to the inchworm algorithm.\cite{Cohen2015}

The 
remainder
of the paper is organized as follows: 
In Sec.~\ref{sec:model}, we define a general impurity model and describe the strong-coupling-expansion formalism based on pseudo-particles.
Section~\ref{sec:sceq} presents our specific way to construct the diagrammatics via self-consistent vertex equations and summarizes the hierarchy of the approximate schemes and the relation to established approaches.
In Sec.~\ref{sec:QdiagMC}, we provide a detailed description of the diagrammatic Monte Carlo algorithm that is used to sample the four-point vertex function.
In Sec.~\ref{sec:Dubiner}, we address the current challenge of the algorithm in storing the four-point vertex function and suggest using a Dubiner basis to compress the data. 
Benchmark results for various types of impurity models are shown in Sec.~\ref{sec:results}.
In Appendix~\ref{app:QwoT}, we present a possible variant of the self-consistent vertex scheme.

\section{Model and Pseudo-particles}\label{sec:model}
We consider a general impurity model whose Hamiltonian can be expressed as 
\begin{align}
	\mathcal{H} &= \mathcal{H}_{\text{loc}} + \mathcal{H}_{\text{hyb}} + \mathcal{H}_{\text{bath}}~,\nonumber\\
	\mathcal{H}_{\text{loc}} &=\sum^{}_{ab}E^{}_{ab}d^{\dagger}_{a}d^{}_{b}+\sum^{}_{abcd}U^{}_{ab,cd}d^{\dagger}_{a}d^{\dagger}_{b}d^{}_{c}d^{}_{d}~,\nonumber\\
	\mathcal{H}_{\text{hyb}} &= \sum_{k\alpha a}\left(V^k_{\alpha a}F^{\dagger}_{\alpha}c^{}_{ka} + \text{h.c.}\right)~,\nonumber\\
	\mathcal{H}_{\text{bath}} &= \sum_{k,ab}\varepsilon^{}_{k,ab}c^{\dagger}_{ka}c^{}_{kb}~,
	\label{eqn:Himp}
\end{align}
where $d^{}_{a}$ ($d^{\dagger}_{a}$) is the annihilation (creation) operator of the impurity fermion with flavor $a$, 
and $c^{}_{ka}$ ($c^{\dagger}_{ka}$) is the annihilation (creation) operator of the bath degree of freedom with index $k$ and flavor $a$.
$F_\alpha$ denotes an operator composed of $d^{\dagger}_{a}$ and $d^{}_{a}$ which depends on the specific model,
and $V^k_{\alpha a}$ the hybridization amplitude. 
For example, this
general form of the impurity model includes the (multi-orbital) Anderson model with fermionic bath degrees of freedom and,
with a truncation to two states (or in the $U\rightarrow \infty$ limit) also 
the spin-boson model with a bosonic bath. 
For the Anderson impurity model, the operator $F^{\dagger}_{\alpha=a}$ is the single fermion operator $d^{\dagger}_a$ and the bath degrees of freedom are fermions, while in the case of the spin-boson model, $F_1$ represents one component of the Pauli spin operator, e.g. $\sum_{ab}d^{\dagger}_{a}\sigma^x_{ab}d^{}_{b}$, and the bath degrees of freedom are bosons. 

We next introduce a pseudo-particle (PP) representation for the diagrammatic treatment of the impurity problem.
For the (many-particle) state $|m\rangle$ in the impurity Hilbert space,
we introduce the pseudo-particle operator $a^{(\dagger)}_{m}$ which connects the pseudo-particle vacuum state $|0\rangle$ to $|m\rangle$ via $|m\rangle = a^{\dagger}_{m}|0\rangle$.
Conventionally, depending on the number of impurity fermions in the state $|m\rangle$, the corresponding pseudo-particle is defined as a boson (even number of fermions) or a fermion (odd number of fermions).\cite{Barnes1976,Coleman1984,Eckstein2010a}
In the pseudo-particle representation, the physical local Hilbert space becomes a subspace of the full Fock space generated by the PP operators, satisfying the constraint
\begin{equation}
	\mathcal{Q}=\sum^{}_{m}a^{\dagger}_{m}a^{}_{m}=1~,
	\label{eqn:Qpp1}
\end{equation}
where $\mathcal{Q}$ is the total pseudo-particle number.
Within the $\mathcal{Q}=1$ subspace, all operators acting on impurity states can be represented via their matrix elements  as quadratic operators in the pseudo-particles. 
For example a general impurity operator can now be expressed as
\begin{align}
	F_{\alpha} = \sum^{}_{mn}v^{\alpha}_{mn}a^{\dagger}_{m}a^{}_n~,
	\label{eqn:ditop}
\end{align}
with $v^\alpha_{mn} = \langle m|F_\alpha|n\rangle$.

After integrating out the bath degrees of freedom, the impurity effective action can be written as 
\begin{equation}
	\mathcal{S}_{\text{imp}} = \mathcal{S}_{\text{loc}} + \mathcal{S}_{\text{hyb}},
	\label{eqn:Seff}
\end{equation}
with the pseudo-particle representations
\begin{align}
	\mathcal{S}_{\text{loc}} =& \sum^{}_{mn}\int_{0}^{\beta}d\tau~h^{}_{mn}a^{\dagger}_{m}(\tau)a^{}_{n}(\tau)~,\\
	\mathcal{S}_{\text{hyb}} =& \sum^{}_{\alpha\gamma}\sum^{}_{mn,m'n'}\int_{0}^{\beta}d\tau\int_{0}^{\beta}d\tau'~a^{\dagger}_{m}(\tau)a^{}_{n}(\tau)\nonumber\\
	&\times (v^\alpha_{nm})^*V_{\alpha\gamma}(\tau-\tau')v^\gamma_{m'n'} a^{\dagger}_{m'}(\tau')a^{}_{n'}(\tau')~\nonumber\\
	=& \sum^{}_{mn,m'n'}\int_{0}^{\beta}d\tau\int_{0}^{\tau}d\tau'~a^{\dagger}_{m}(\tau)a^{}_{n}(\tau)\nonumber\\
	&\times \overline{V}^{mn}_{m'n'}(\tau-\tau') a^{\dagger}_{m'}(\tau')a^{}_{n'}(\tau')~,
	\label{eqn:Seff_terms}
\end{align}
and $\beta$ the inverse temperature $1/T$.
The local action is now expressed with a quadratic potential $h_{mn}=\langle m|\mathcal{H}_{\text{loc}}|n\rangle$, while the hybridization term becomes a retarded interaction between pseudo-particles. The Matsubara-frequency expression for the retarded interaction is
$V_{\alpha\gamma}(i\omega_n) = \sum_{k,ab}V^{k}_{\alpha a}\left[i\omega_n-\varepsilon_k\right]^{-1}_{ab}(V^{k}_{\gamma b})^*$. 
Summing over $\alpha$ and $\gamma$ one may then absorb the matrix elements into the pseudo-particle interaction $\overline{V}^{mn}_{m'n'}$.

The physical Hamiltonian conserves the total pseudo-particle number by construction. One way to impose the constraint  \eqref{eqn:Qpp1} is by extracting the leading terms in the low-density expansion within a grand canonical formulation with respect to $\mathcal{Q}$.\cite{Barnes1976,Coleman1984} This results in a single, directed pseudo-particle backbone line dressed by hybridization lines. Alternatively, the same diagrammatic equations are obtained by expressing the direct Taylor expansion of the partition function and observables with respect to $\mathcal{S}_{\rm hyb} $ in terms of time-ordered expectation values (resolvent operators).\cite{Keiter1971,Aoki2014} In the following we will summarize these equations (the derivation has been been given in the literature), and then explain  in more detail how to resum them in terms of the three-point and four-point vertex.

The constrained pseudo-particle propagator associated with the 
action \eqref{eqn:Seff}
is defined for $0\le \tau\le \beta$ as 
\begin{equation}
	\mathcal{G}_{mn}(\tau)=\left\langle a_{m}(\tau)a^{\dagger}_n(0)\right\rangle^{\mathcal{Q}=1}_{\mathcal{S}_{\text{imp}}} \, ,
	\label{eqn:Gpp}
\end{equation}
and the corresponding pseudo-particle self-energy via the 
time-ordered
Dyson equation
\begin{align}
	&\mathcal{G}_{mm'}(\tau)=\mathcal{G}_{0,mm'}(\tau) + \sum^{}_{m_1m_2}\int_{0}^{\tau}d\tau_1\int_{\tau_1}^{\tau}d\tau_2\nonumber\\
	&\times\mathcal{G}_{mm_1}(\tau-\tau_2)\Sigma_{m_1m_2}(\tau_2-\tau_1)\mathcal{G}_{0,m_2m'}(\tau_1)~.
	\label{eqn:ppDyson}
\end{align}
Here $\mathcal{G}_{0,mm'}(\tau)=\langle a^{}_{m}(\tau)a^{\dagger}_{m'}(0)\rangle_{\mathcal{S}_{\text{loc}}}^{\mathcal{Q}=1}$, and we have the time ordering $0\le \tau_1\le \tau_2 \le \tau \le \beta$.  Depending on the particle statistics, the PP propagators with negative time argument are defined as $\mathcal{G}_{mn}(-\tau)=\xi_m\mathcal{G}_{mn}(\beta-\tau)$, where $\xi_m=1$ for bosonic PP and $-1$ for fermionic PP, respectively.
\cite{fermionicPPrep}  

By definition, we have the boundary condition $\mathcal G(0_+)=\mathcal G_0(0_+)=\mathcal I$, with $\mathcal I$ the identity matrix in PP space.  The restriction of $\mathcal G$ to the subspace $\mathcal{Q}=1$ implies $\sum_{m} \mathcal{G}_{mm}(\beta_-)= \sum_{m} \xi_m\mathcal{G}_{mm}(0_-)=1$. To satisfy this constraint, we introduce an auxiliary pseudo-particle chemical potential shift $\lambda$, as described below.

\section{Self-consistency equations}\label{sec:sceq}

\begin{figure}[t]
	\centering
	\includegraphics[height=0.45\textwidth,angle=270]{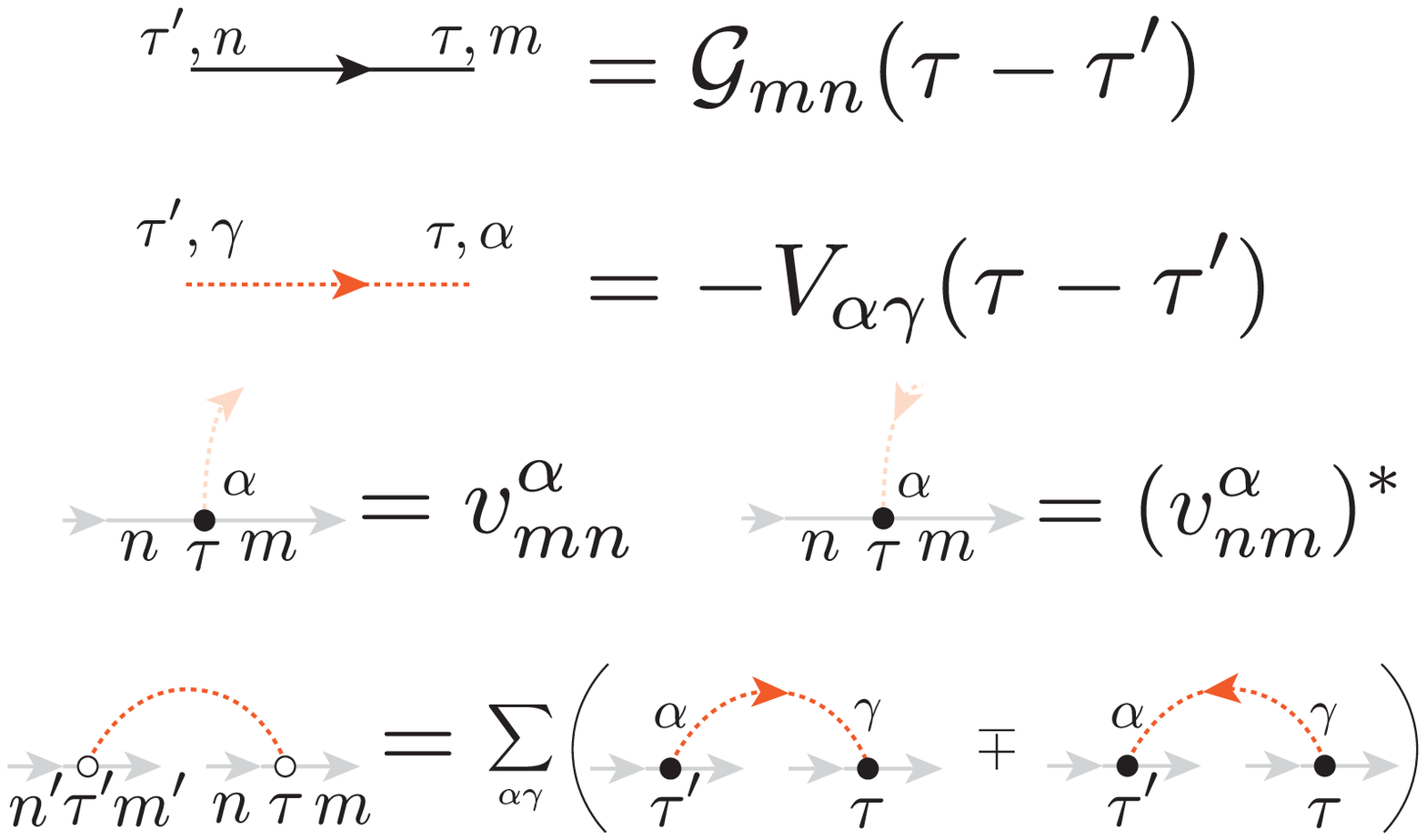}
	\caption{
		Building blocks of the PP diagrams. From top to bottom, the (black) solid line, (red) dashed line, and the (black) solid circle represent the PP propagator, the retarded interaction, and the (bare) vertex.
		The bottommost row shows the undirected retarded interaction introduced to simplify the diagrams.
		The minus (plus) sign in the parentheses is for fermionic (bosonic) bath degrees of freedom.
	}
	\label{fig:block}
\end{figure}

\begin{figure}[t]
	\centering
	\includegraphics[width=0.45\textwidth]{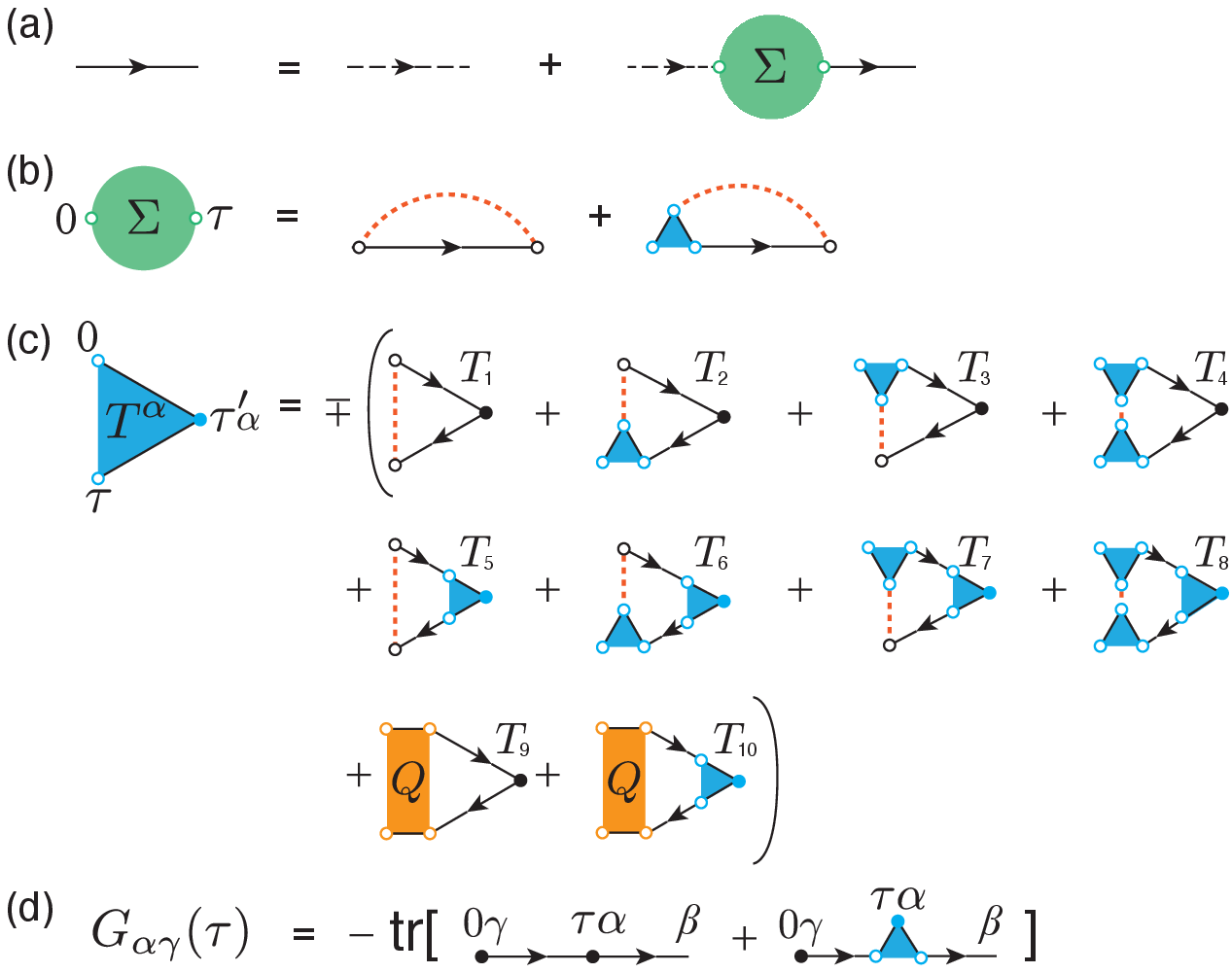}
	\caption{
		Diagrammatic representation of the self-consistency equations.
		(a) Dyson equation for the PP propagator (Eq.~\eqref{eqn:ppDyson}) in terms of the PP self-energy $\Sigma$ and the bare PP propagator [black dashed arrow]. (b) PP self-energy as a function of the triangular vertex $T$ (Eq.~\eqref{eqn:SigInGVT}). (c) Vertex self-consistency equation which involves the four-point vertex $Q$. (d) Impurity Green's function (Eq.~\eqref{eqn:Gimp}).
		The global sign in panel (c) is determined by the particle statistics: fermion ($-$) and boson (+).
	}
	\label{fig:scEqn}
\end{figure}

In this study, we use skeleton diagrams to compute the PP self-energy. 
The building blocks of the PP diagrams, illustrated in Fig.~\ref{fig:block}, are the renormalized propagator $\mathcal{G}_{mn}(\tau-\tau')$, the retarded interaction $V_{\alpha\gamma}(\tau-\tau')$ (directed interaction lines), and the interaction vertices $v^\alpha_{mn}$ and $\left(v^\alpha_{nm}\right)^*$. The $n$th-order PP self-energy $\Sigma_{mn}(\tau)$ consists of a sequence of $2n-1$ renormalized propagator lines (``backbone''). $n$ interaction lines connect the $2n$ vertices on the backbone line in such a way that the backbone line is one-particle irreducible (1PI), i.e, it cannot be separated by cutting the backbone. Note that there is no PP loop beside the backbone line due to the $\mathcal{Q}=1$ constraint. In the case of a fermion bath  
and fermionic coupling operators $F$, the diagram acquires an additional sign  $(-1)^{C}$, where $C$ is the sum of the number of interaction lines which are directed against the backbone, plus the number of line crossings.  Since the series includes all possible directions of the interaction lines, it is convenient to introduce the undirected retarded interaction $\overline{V}^{mn}_{m'n'}(\tau-\tau')$ (see Eq.~\eqref{eqn:Seff_terms}), which  combines the forward and backward hybridization line with the vertices, and includes also the fermion sign associated with the direction of the line (see Fig.~\ref{fig:block}, lower panel).
In the skeleton diagram technique, the propagators in the backbone of the self-energy diagram are self-consistently determined via Eq.~(\ref{eqn:ppDyson}); see also Fig.~\ref{fig:scEqn}(a).


In addition to the self-consistent calculation of the renormalized PP propagator, we also self-consistently determine the triangular vertex.
All possible skeleton diagrams of the PP self-energy can be expressed in a compact form using the triangular vertex $\overline{T}^{m_1n_1}_{m_2n}(\tau_2,\tau_1)$: 
\begin{align}
	\Sigma_{mn}(\tau) =& \sum_{m'n'}
	\mathcal{G}_{m'n'}(\tau)
	\overline{V}^{mm'}_{n'n}(\tau) + \sum_{\substack{m_1n_1\\m_2n_2}}\int_{0}^{\tau}d\tau_1\int_{\tau_1}^{\tau}d\tau_2\nonumber\\
	&\times\mathcal{G}_{n_2m_2}(\tau-\tau_2)\overline{V}^{mn_2}_{m_1n_1}(\tau-\tau_1)\overline{T}^{m_1n_1}_{m_2n}(\tau_2,\tau_1)~.
	\label{eqn:SigInGVT}
\end{align}
Figure~\ref{fig:scEqn}(b) shows the diagrammatic representation of Eq.~(\ref{eqn:SigInGVT}).
The triangular vertex $\overline{T}$ defines a vertex $T^\alpha$ with specified operator index $\alpha$ through $T^{\alpha}_{mn} = \sum^{}_{m_1n_1}v^{\alpha}_{m_1n_1}\overline{T}^{m_1n_1}_{mn}$. As illustrated in Fig.~\ref{fig:scEqn}(c), $T^\alpha$ can be determined by a self-consistency equation, 
which
involves a four-point vertex (orange box). 
This equation
will be discussed in detail in Sec.~\ref{ssec:vsc}.
Finally, using the self-consistent PP 
propagators
and the triangular vertex $T^\alpha$, one can compute the impurity Green's function $G_{\alpha\gamma}(\tau)$ via
\begin{align}
	G_{\alpha\gamma}(\tau) =& -\langle\mathcal{T}_\tau F^{}_{\alpha}(\tau)F^{\dagger}_{\gamma}(0)\rangle\nonumber\\
	=& -\tr\left[\mathcal{G}(\beta-\tau)v^\alpha\mathcal{G}(\tau)\left(v^\gamma\right)^{\dagger}\right] - \int_{0}^{\tau}d\tau_1\int_{\tau}^{\beta}d\tau_2~\nonumber\\
	&\times \tr\left[\mathcal{G}(\beta-\tau_2)T^\alpha(\tau_2-\tau_1,\tau-\tau_1)\mathcal{G}(\tau_1)\left(v^\gamma\right)^{\dagger}\right]~,
	\label{eqn:Gimp}
\end{align}
see illustration in Fig.~\ref{fig:scEqn}(d).

\subsection{Vertex self-consistency equation}\label{ssec:vsc}
\begin{figure}[t]
	\centering
	\includegraphics[height=0.45\textwidth,angle=270]{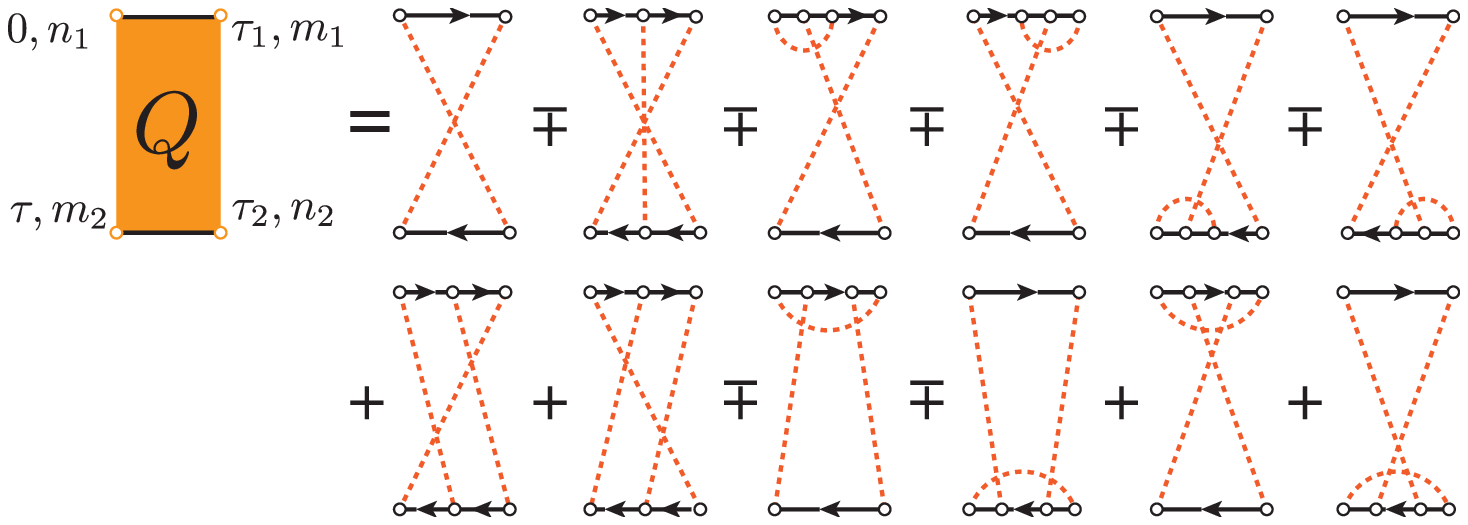}
	\caption{
		Expansion of the $Q$ vertex up to the third order. There is a single second-order diagram, and there are eleven third-order contributions.
		The upper (lower) sign on the right-hand side corresponds to fermions (bosons).
	}
	\label{fig:Qlow}
\end{figure}

Figure~\ref{fig:scEqn}(c) shows the diagrammatic representation of the vertex self-consistency equation.
The triangular vertex 
$T^\alpha$
in Fig.~\ref{fig:scEqn}(c) can be written as a sum of diagrams which involve the renormalized PP Green's function, the retarded (bare) interaction $\overline{V}$, $T^\alpha$ 
itself, and in addition the PP four-point vertex $Q$.
$Q^{m_1n_1}_{m_2n_2}(\tau;\tau_2,\tau_1)$ is a diagrammatic object with two separate renormalized propagator sequences (two backbones) connected by  retarded interactions.
Figure~\ref{fig:Qlow} illustrates the twelve lowest-order $Q$ diagrams (up to order $3$).
Note that we exclude diagrams with a single transverse interaction line between the two backbones, and two-particle reducible diagrams in the propagator line, to avoid a double counting with the first eight diagrams in Fig.~\ref{fig:scEqn}(c).

In Fig.~\ref{fig:scEqn}(c), 
due to the different time-ordering rule depending on the particle statistics, the overall sign on the right-hand side differs for fermions ($-$) and for bosons (+).
This rule ensures that for a given self-energy diagram in Fig.~\ref{fig:scEqn}(b) which is obtained by expanding the vertex into bare PP Green's functions and undirected interaction lines, the sign is  $\pm1$ for an even (odd) number of crossings of interaction lines for fermions.
Note that an overall $(-1)^n$ expansion factor for diagram order $n$ and the signs associated with the time orderings between 
the creation and annihilation operators attached to the
hybridization lines are absorbed into the dashed lines. 

Due to the lack of bubbles in the triangular vertex diagram, the self-consistency equation is causal in imaginary time, and
the integro-differential equation represented by Figs.~\ref{fig:scEqn}(a) and (c) can be efficiently solved by a time-stepping procedure.
In practice we transform Eq.~(\ref{eqn:ppDyson}), corresponding  to Fig.~\ref{fig:scEqn}(a), into an integral-differential form suitable for time stepping,
\begin{equation}
	\left[-\partial_{\tau} - h  + \lambda\right]\mathcal{G}(\tau) + \int_{0}^{\tau}d\tau'~\Sigma(\tau-\tau')\mathcal{G}(\tau')=0~, 
	\label{eqn:Volterra}
\end{equation}
with $0\le \tau\le \beta$. Here, we omit the pseudoparticle indices. 

\begin{figure}[]
	\centering
	\includegraphics[height=0.45\textwidth,angle=270]{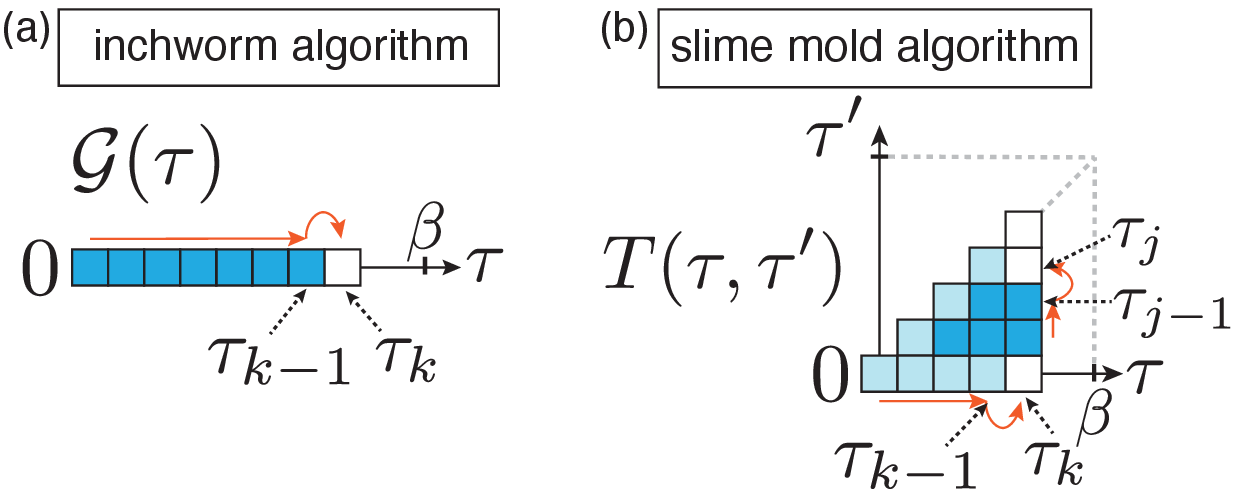}
	\caption{
		Schematic picture showing the time-stepping procedure of (a) the inchworm algorithm and (b) the slime mold algorithm. 
		(Red) arrows present the time-stepping direction in the inchworm and slime mold algorithm.
		In the slime mold algorithm, the bulk (dark blue) and boundary (light blue) time points are treated separately (see sec.~\ref{ssec:vsc}); for a given time slice $\tau_k$, the bulk time points ($0<\tau_j<\tau_k$) are updated using Eq.~\eqref{eqn:vscBulk} before the boundary points ($\tau_j=0$ and $\tau_k$) are updated using Eq.~\eqref{eqn:vscBoundary}. In practice, we update the bulk time points in a time-ascending order, although one can in principle freely choose the order.
	}
	\label{fig:worm_vs_slime}
\end{figure}

The time-stepping solution for the vertex self-consistency in Fig.~\ref{fig:scEqn}(c) can be performed as follows:
For the numerical solution, the imaginary-time interval $[0,\beta]$ is discretized into $N$ time slices of length $\Delta\tau=\beta/N$, which defines the $N+1$ imaginary-time points $\tau_i=i\Delta\tau$ ($i=0,1,\ldots,N$). The time arguments of the vertex $T^\alpha(\tau_i,\tau_j)$ satisfy $\beta \ge \tau_i\ge \tau_j\ge 0$, and after $k$ steps, the vertex is known for all $\tau_k\ge \tau_i\ge \tau_j\ge 0$. In the next step of the procedure, $T^\alpha(\tau_{k+1},\tau_j)$ is calculated using the previously computed $T^\alpha(\tau_i,\tau_j)$ values up  to $\tau_k\ge \tau_i\ge \tau_j\ge 0$.
Starting from $T^\alpha(0,0)$, one may thus extend the solution time-step by time-step up to the maximum time $\tau_N=\beta$.
Such a procedure can be considered as a two-dimensional generalization of the inchworm algorithm,\cite{Cohen2015} see the illustration in  Fig.~\ref{fig:worm_vs_slime}. 
Inspired by the two-dimensional propagation of the slime mold creature ({\it physarum polycephalum}),\cite{Slime2016} we may thus refer to the vertex-based algorithm as a slime mold algorithm.

For a given maximum time index $k$ and $i\geq j \geq 0$, the triangular vertex self-consistency equation can either be solved via matrix inversion or in an iterative manner.
For the bulk case, $k>j>0$,
\begin{equation}
	T^{\alpha}_{mn}(\tau_k,\tau_j) = B^{\alpha}_{mn}(\tau_k,\tau_j) + \sum^{}_{m'n'}M^{m'n}_{mn'}(\tau_k,\tau_j)T^\alpha_{n'm'}(\tau_k,\tau_j), 
	\label{eqn:vscBulk}
\end{equation}
while for the boundary case, $j=0$ or $j=k$,
\begin{align}
	T^{\alpha}_{mn}(\tau_k,\tau_j) &= \overline{B}^{\alpha}_{mn}(\tau_k,\tau_j)\nonumber\\
	&+ \sum_{\gamma,m'n'}\left[\overline{M}^{m'n}_{mn'}\right]^{\alpha\gamma}(\tau_k,\tau_j)T^{\gamma}_{n'm'}(\tau_k,\tau_j)\nonumber\\
	&+ \sum_{\gamma,m'n'}\left[\overline{L}^{m'n}_{mn'}\right]^{\alpha\gamma}(\tau_k,\overline{\tau}_j)T^{\gamma}_{n'm'}(\tau_k,\overline{\tau}_j),
	\label{eqn:vscBoundary}
\end{align}
in which $\overline{\tau}_j=0$ ($\tau_k$) when $\tau_j=\tau_k$ ($0$).
Here, $B$  and $\overline{B}$ represent the terms that do not include $T(\tau_i,\tau_j)$ contributions with 
$i=k$
and $M$, $\overline{M}$, and $\overline{L}$ are the coefficients of $T$-linear contributions with the bigger time index $i=k$.
Since the zero-range time integral vanishes, only $T_5$ and $T_{10}$ ($T_2$ and $T_3$) contribute to $M$ ($\overline{M}$ and $\overline{L}$).

Alternatively, one can also start from an initial guess for $T$ (e.g. $T=T_1$) and continue to update $T$ by substituting the current guess into the right-hand side of the self-consistency equation until convergence. 
In practice, we apply this iterative scheme to obtain the boundary values within the time-stepping procedure (instead of solving Eq.~(\ref{eqn:vscBoundary})), while for the bulk values we perform the time stepping, as indicated by the vertical red arrows in Fig.~\ref{fig:worm_vs_slime}(b).
It turns out that in this way the convergence of the iterative method is fast and typically requires only of the order of 10 iterations for the presented parameters.

\subsection{Approximation hierarchy}

In this subsection, we relate our diagrammatic formulation to different perturbative approximation schemes.
In NCA,\cite{Keiter1971} the self-energy is approximated by the first-order contribution, i.e., the first term in Fig.~\ref{fig:scEqn}(b), and vertex corrections are neglected.
The one-crossing approximation (OCA)\cite{Pruschke1989} takes into account the lowest-order vertex correction, diagram $T_1$ in Fig.~\ref{fig:scEqn}(c), which leads to a second-order self-energy in the skeleton series.
The name two-crossing approximation (TCA) will be used to refer to the approximation which further includes the vertex corrections $T_2$, $T_3$, and $T_5$ in which the renormalized vertex (blue triangle) is substituted by the OCA vertex.
Note that in NCA, OCA, and TCA, the self-consistency for the triangular vertex is not applied.
We also note that the TCA scheme does not include the second-order vertex diagram contributed by $T_9$  (with $Q$ replaced by the second order diagram in Fig.~\ref{fig:Qlow}), which gives rise to a contribution of third order in the interaction to the self-energy. The approximation containing all third order diagrams for the self-energy will be called  the third-order approximation (TOA).~\cite{Eckstein2010}

The triangular vertex approximation (TVA) is the lowest-order approximation in this study that introduces the vertex self-consistency [Fig.~\ref{fig:scEqn}(c)], but without the $Q$ vertex contributions (diagrams $T_9$, $T_{10}$).
To further improve on this, we incorporate the four-point vertex by successively increasing the diagram order of $Q$.
The self-consistent vertex scheme with $n$th-order $Q$ vertex is denoted by $Q_n$.
For example, $Q_3$ takes into account the four-point vertex diagrams shown in Fig.~\ref{fig:Qlow} within the fully self-consistent scheme of Fig.~\ref{fig:scEqn}(c).
$Q_n$ converges to the exact four-point vertex in the $n\rightarrow\infty$ limit if the series representation of $Q$ is convergent.

\section{Monte Carlo sampling of the Four-Point Vertex} \label{sec:QdiagMC}
\subsection{diagMC formalism}
We sample the Feynman diagrams of the four-point vertex $Q^{m_1n_1}_{m_2n_2}(\tau;\tau_2,\tau_1)$ using a bold-line diagrammatic Monte Carlo (diagMC)~\cite{Prokofev2008,Prokofev2007} method based on the strong-coupling expansion.\cite{Aaram2021}
The series representation of $Q^{m_1n_1}_{m_2n_2}(\tau;\tau_2,\tau_1)$ is $Q^{m_1n_1}_{m_2n_2}(\tau;\tau_2,\tau_1) = \sum^{}_{n}\left[Q^{(n)}\right]^{m_1n_1}_{m_2n_2}(\tau;\tau_2,\tau_1)$ with
\begin{widetext}
\begin{eqnarray}
\left[Q^{(n)}\right]^{m_1n_1}_{m_2n_2}(\tau;\tau_2,\tau_1) &=& \sum_{X\in\mathrm{topology}}\sum^{}_{\{\bm{r}^u_i\},\{\bm{r}^d_i\}}\delta(\tau_0^u)\delta(\tau^u_{k_u+1}-\tau_1)\delta(\tau^d_{0}-\tau_2)\delta(\tau^d_{k_d+1}-\tau)~\omega^{m_1n_1}_{m_2n_2}\left(\{X,\{\bm{r}^u_i\},\{\bm{r}^d_i\}\}\right),\hspace{6mm}\label{eqn:Qn}\\
	\omega^{m_1n_1}_{m_2n_2}(\{X,\{\bm{r}^u_i\},\{\bm{r}^d_i\}\}) &=& \Pi_{X}(\{\bm{r}^u_i\},\{\bm{r}^d_i\})\nonumber\\
	&&\times\left[{v}(\bm{r}^u_{k_u+1}){\mathcal{G}}(\tau^u_{k_u+1}-\tau^u_{k_u})\dots {v}(\bm{r}^u_{1}){\mathcal{G}}(\tau^u_{1}-\tau^u_{0}){v}(\bm{r}^u_{0})\right]_{m_1n_1}\nonumber\\
	&&\times\left[{v}(\bm{r}^d_{k_d+1}){\mathcal{G}}(\tau^d_{k_d+1}-\tau^d_{k_d})\dots {v}(\bm{r}^d_{1}){\mathcal{G}}(\tau^d_{1}-\tau^d_{0}){v}(\bm{r}^d_{0})\right]_{m_2n_2}~,\label{eqn:wMC}
\end{eqnarray}
\end{widetext}
where $\delta(\tau)$ is the Dirac delta function.
In Eqs.~(\ref{eqn:Qn}) and (\ref{eqn:wMC}), we introduced the combined coordinate for the upper ($u$) or lower ($d$) backbone, $\bm{r}^{u/d}=(\alpha,t,\tau)$, where $\alpha$ is the $F$ operator index, $t=\pm$ refers to the type of operator (creation/annihilation), and $\tau$ is the imaginary time. For example, the vertex 
${v}(\bm{r}=(\alpha,-,\tau))={v}^\alpha$, while ${v}(\bm{r}=(\alpha,+,\tau))=\left[{v}^\alpha\right]^{\dagger}$, both of them representing matrices in pseudo-particle space. 
The sum over combined coordinates in Eq.~(\ref{eqn:Qn}) represents the sum over the types and flavors, and the time-ordered integrals:
\begin{align}
	\sum^{}_{\{\bm{r}^u_i\},\{\bm{r}^d_i\}} =& \sum^{2n-4}_{\substack{k_u,k_d=0\\k_u+k_d=2n-4}}\sum^{N_F}_{\alpha^{u}_i,\alpha^{d}_i=1}\sum^{}_{t^u_i,t^d_i=\pm}\nonumber\\
	&\times\int_{0}^{\tau_1}d\tau^u_{0}\int_{\tau^u_0}^{\tau_1}d\tau^u_{1}\dots\int_{\tau^u_{u_{k}-1}}^{\tau_1}d\tau^u_{{k_u}}\int_{\tau^u_{k_u}}^{\tau_1}d\tau^u_{{k_u+1}}\nonumber\\
	&\times\int_{\tau_2}^{\tau}d\tau^d_{0}\int_{\tau^d_0}^{\tau}d\tau^d_{1}\dots\int_{\tau^d_{{k_d-1}}}^{\tau}d\tau^d_{{k_d}}\int_{\tau^d_{k_d}}^{\tau}d\tau^d_{{k_d+1}}~.
	\label{eqn:primedSum}
\end{align}
$\Pi_{X}(\{\bm{r}^u_i\},\{\bm{r}^d_i\})$ in Eq.~(\ref{eqn:wMC}) represents the product of the interaction lines for a given topology of connections $X$ and the vertex configuration  $\{\bm{r}^u_i\},\{\bm{r}^d_i\}$.

\begin{figure}[t]
	\centering
	\includegraphics[height=0.45\textwidth,angle=270]{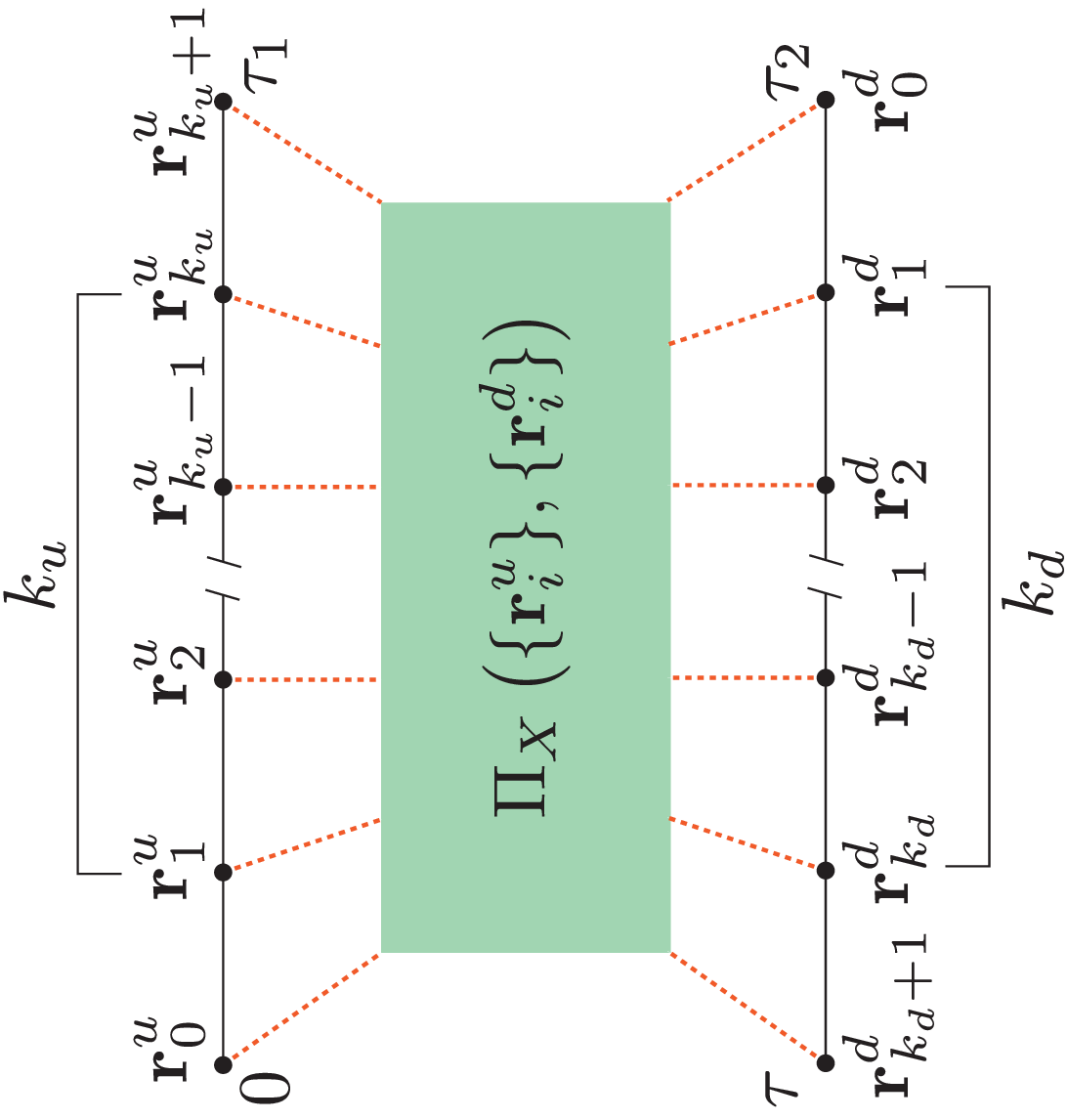}
	\caption{
		Monte Carlo configuration composed of a set of upper ($\{v^u_{i}\}$) and lower ($\{v^d_{i}\}$) vertices connected by interaction lines of topology $X$.
		The numbers of the upper and lower (inner) vertices are $k_u$ and $k_d$, respectively.
	}
	\label{fig:QnNotation}
\end{figure}

The configuration space of the MC sampling, $\{ \mathbf{x} \}=\left\{X,\{\bm{r}^u_i\},\{\bm{r}^d_i\}\right\}$, 
is composed of the topological structure of the interaction lines $X$, and the set of vertices on the upper and lower backbone.
The weight of these configurations, which are illustrated in Fig.~\ref{fig:QnNotation}, is given by Eq.~(\ref{eqn:wMC}).
Note that only configurations with $X$ corresponding to 2-particle irreducible (2PI) diagrams along the PP propagators and 1PI diagrams along the interaction lines are included.
Here, the 2PI condition is examined with the external vertices being connected by auxiliary backbone lines, as shown by the gray lines in Fig.~\ref{fig:MC_insertion_removal}.
Both this 2PI condition and the 1PI condition for the interaction lines are needed to prevent the double-counting of $T$ diagrams in Fig.~\ref{fig:scEqn}.

In practice we use a combined MC weight for the configurations with different external subspace indices, defined via the $L^1$ norm $|\omega_c(\mathbf{x})| = \sum^{}_{m_1n_1,m_2n_2}|\omega^{m_1n_1}_{m_2n_2}(\mathbf{x})|$, and measure the $\left[Q^{(n)}\right]^{m_1n_1}_{m_2n_2}$ component by accumulating $\omega^{m_1n_1}_{m_2n_2}/|\omega_c|$.

\subsection{Monte Carlo updates}
In order to ensure the ergodicity of the Monte Carlo sampling, we use four different updates: (1) insertion, (2) removal, (3) swap, and (4) shift line-cut. 
These updates will be briefly explained in the following. 
\subsubsection{Insertion and removal update}
\begin{figure}[h]
	\centering
	\includegraphics[height=0.40\textwidth,angle=270]{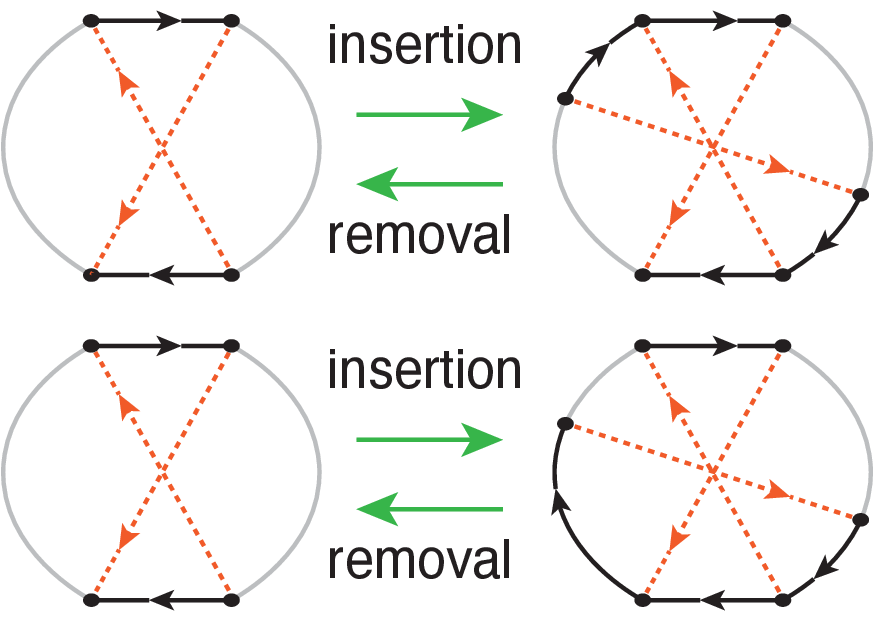}
	\caption{Examples of insertion and removal updates between orders 2 and 3, which involve the fictitious propagators (gray lines).}
	\label{fig:MC_insertion_removal}
\end{figure}
In an insertion update, we randomly select two distinct 
PP propagators and place the two end points of the
new
 interaction line $V$ 
on those lines.
For this step, we close the diagram in Fig.~\ref{fig:QnNotation} into a circle by inserting fictitious 
PP lines
that connect the end points $(0,\tau)$ and $(\tau_1,\tau_2)$ of the $Q$ diagram; see gray lines in Fig.~\ref{fig:MC_insertion_removal}. 
When an operator is inserted on a fictitious line, it defines a new corner of the $Q$ diagram, and we have a choice as to which of the two backbone branches will be extended to the new operator. Two possible choices are illustrated in Fig.~\ref{fig:MC_insertion_removal}. The proposal probability of the insertion update has to take into account this degree of freedom. If the move is accepted,  
either one of the external time points ($\tau,\tau_2,\tau_1$) is shifted,  or, if the first operator on the upper branch is modified, it defines the new $\tau=0$ point and all the other time points in the diagram are shifted accordingly. 

Depending on the nonzero components of the retarded interaction, the inserted vertices $v$, which are understood here as matrices in pseudo-particle space, are chosen differently. 
During the MC insertion update, we only propose combinations of vertices with nonzero retarded interaction.
For a general 
$V_{\alpha\gamma}$ with off-diagonal 
components, there are $2N_F^2$ ($N_F$ is the number of $F$ operator indices) possible combinations of operators.
Here, the factor $2$ comes from the direction of the $V$ line, which determines the location of the creation and annihilation operator.

If the diagram before the insertion was 2PI, the updated diagram automatically satisfies the 2PI condition. Hence, a topology check is not required for the insertion update.

In the removal update, we randomly remove one of the $V$ lines.
Since the topology checking is the most expensive routine, the 2PI condition is examined only in the case of acceptance in the Metropolis step.

The proposal probabilities of the insertion ($W_i$) and removal ($W_r$) updates between $k$th-order and $(k+1)$th-order diagrams are
\begin{align}
	\begin{split}
		W_i &= \left(\frac{1}{2}\right)^{N_{\mathrm{fictitious}}}\frac{2}{(2k)(2k-1)}\frac{1}{2N_F^2}\frac{d\tau^2}{\Delta\tau\Delta\tau'}~,\\
		W_r &= \frac{1}{k+1}~,
		\label{eqn:propProb}
	\end{split}
\end{align}
where $N_{\mathrm{fictitious}}\in\{0,1,2\}$ denotes the number of selected fictitious lines, and $\Delta\tau$, $\Delta\tau'$ represent the lengths of the two PP propagator lines on the imaginary-time axis. 

The resulting acceptance probability for an insertion update from configuration $\mathbf{x}$ to $\mathbf{y}$ is $\text{min}(1,R_i(\mathbf{x}\rightarrow\mathbf{y}))$ where
\begin{equation}
	R_i(\mathbf{x}\rightarrow\mathbf{y}) = \frac{2^{N_{\mathrm{fictitious}}} (2k)(2k-1) N_F^2 \Delta\tau\Delta\tau'}{k+1}\frac{|\omega(\mathbf{y})|}{|\omega(\mathbf{x})|}~.
	\label{eqn:accProb}
\end{equation}

\subsubsection{Swap update}
\begin{figure}[t]
	\centering
	\includegraphics[height=0.40\textwidth,angle=270]{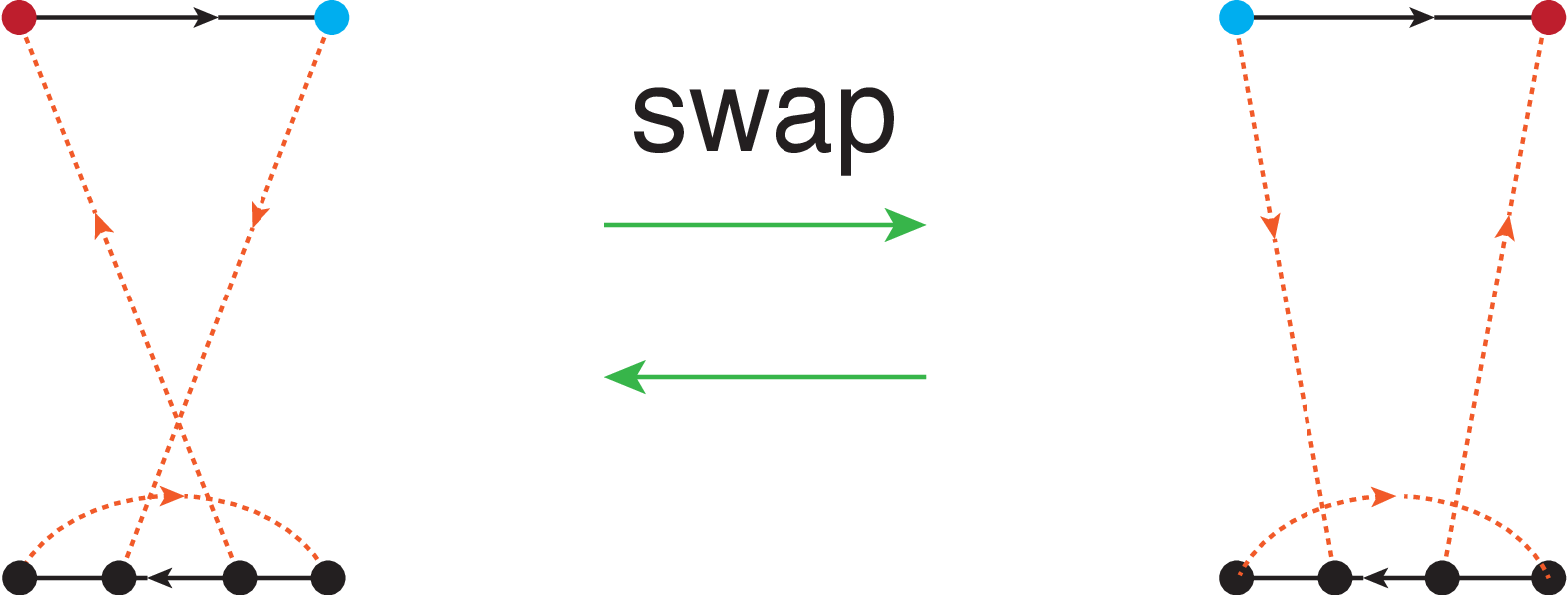}
	\caption{
		Example of a swap update in a third order diagram.
	}
	\label{fig:MC_swap}
\end{figure}

In a swap update, we randomly choose two $V$ lines and swap the end-points of the lines 
with the corresponding fermionic operators, see illustration in Fig.~\ref{fig:MC_swap}. 
This update is essential to ensure an ergodic sampling of the third-order diagrams.
When the outgoing external vertex of the upper backbone and the incoming one of the lower backbone are the same,\footnote{Without this condition, the right diagram could be generated via insertion and shift line-cut updates.} 
for example, one cannot access the right diagram of Fig.~\ref{fig:MC_swap} through an insertion update, since none of the possible second-order diagrams are 2PI.

We also check the 2PI condition after the swap update.

\subsubsection{Shift line-cut}
In this update, we swap the fictitious line connecting $\tau_1$ and $\tau_2$ with an existing PP propagator line, 
thereby relocating the edges $\tau_1$ and $\tau_2$ of the $Q$ vertex. 
The 2PI condition has to be checked after this update as well.

\subsection{Measurements}
At every MC step, we accumulate $\omega^{m_1n_1}_{m_2n_2}(\tau,\tau_2,\tau_1;\mathbf{x})/|\omega_c(\tau,\tau_2,\tau_2;\mathbf{x})|$ for all non-zero $(m_1n_1,m_2n_2)$ combinations.

\subsection{Normalization}
In addition to the vertex function $Q$ we also measure 
observables proportional to the absolute value integral 
\begin{align}
	\mathcal{N}^{(n)} =& \sum^{}_{m_1n_1,m_2n_2}\int_{0}^{\beta}d\tau\int_{0}^{\tau}d\tau_2\int_{0}^{\tau_2}d\tau_1~\nonumber\\
&\times\sum_{\mathbf{x}\in (\text{order $n$})}\left\vert\omega^{m_1n_1}_{m_2n_2}(\tau,\tau_2,\tau_1;\mathbf{x})\right\vert
	\label{eqn:normalization}
\end{align}
during the Monte Carlo sampling. 
Using the analytically computed $\mathcal{N}^{(2)}$, or the previously sampled $\mathcal{N}^{(n-1)}$, we normalize the accumulated $Q^{(n)}$ to obtain the final result.
\begin{figure}[]
	\centering
	\includegraphics[width=0.5\textwidth]{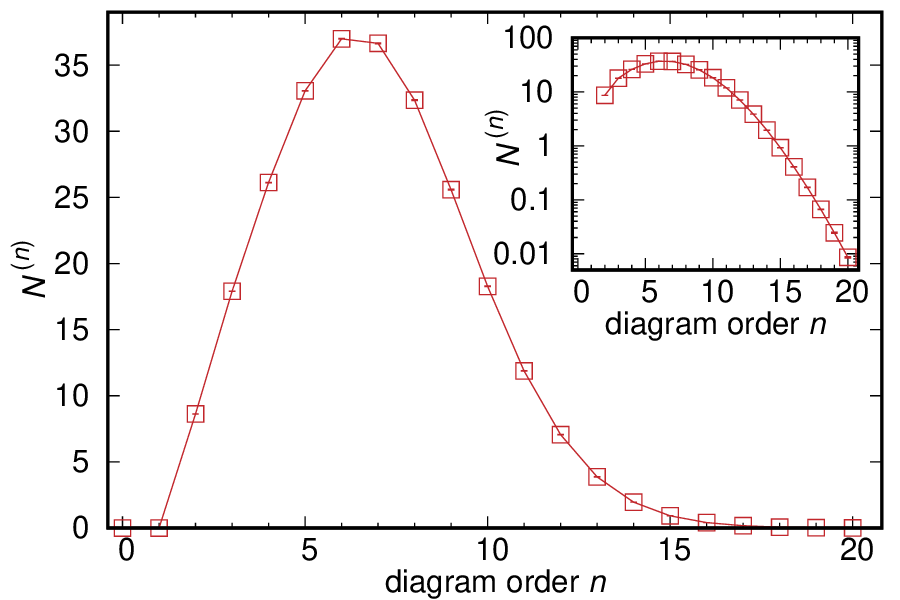}
	\caption{
		Absolute value integral of the $Q$ vertex [Eq.~(\ref{eqn:normalization})] for the single-orbital impurity model (Sec.~\ref{sec:benchmarkSingleBath}) with $T=0.1$, $U=1$, $\mu=0.5$, $V=0.5$ and $\varepsilon=0$.
		The inset presents the same data on a semilog scale, which shows a linear decay for $n\gtrsim 15$.
	}
	\label{fig:normalization}
\end{figure}
As an illustration, Fig.~\ref{fig:normalization} shows the simulation results for the absolute value integral defined in Eq.~(\ref{eqn:normalization}) for a single-orbital impurity model with $T=U/10$, $\mu=U/2$, constant hybridization $V=U/2$, 
and $\varepsilon=0$ (see sec.~\ref{sec:benchmarkSingleBath}).
The integral has a clear peak around order 6 and exponentially decreases as we further increase the diagram order.
In the case of a small average sign, the Monte Carlo error is directly proportional to the absolute-value integral.

\subsection{2-particle irreducibility}

\begin{figure}[t]
	\centering
	\includegraphics[width=0.4\textwidth]{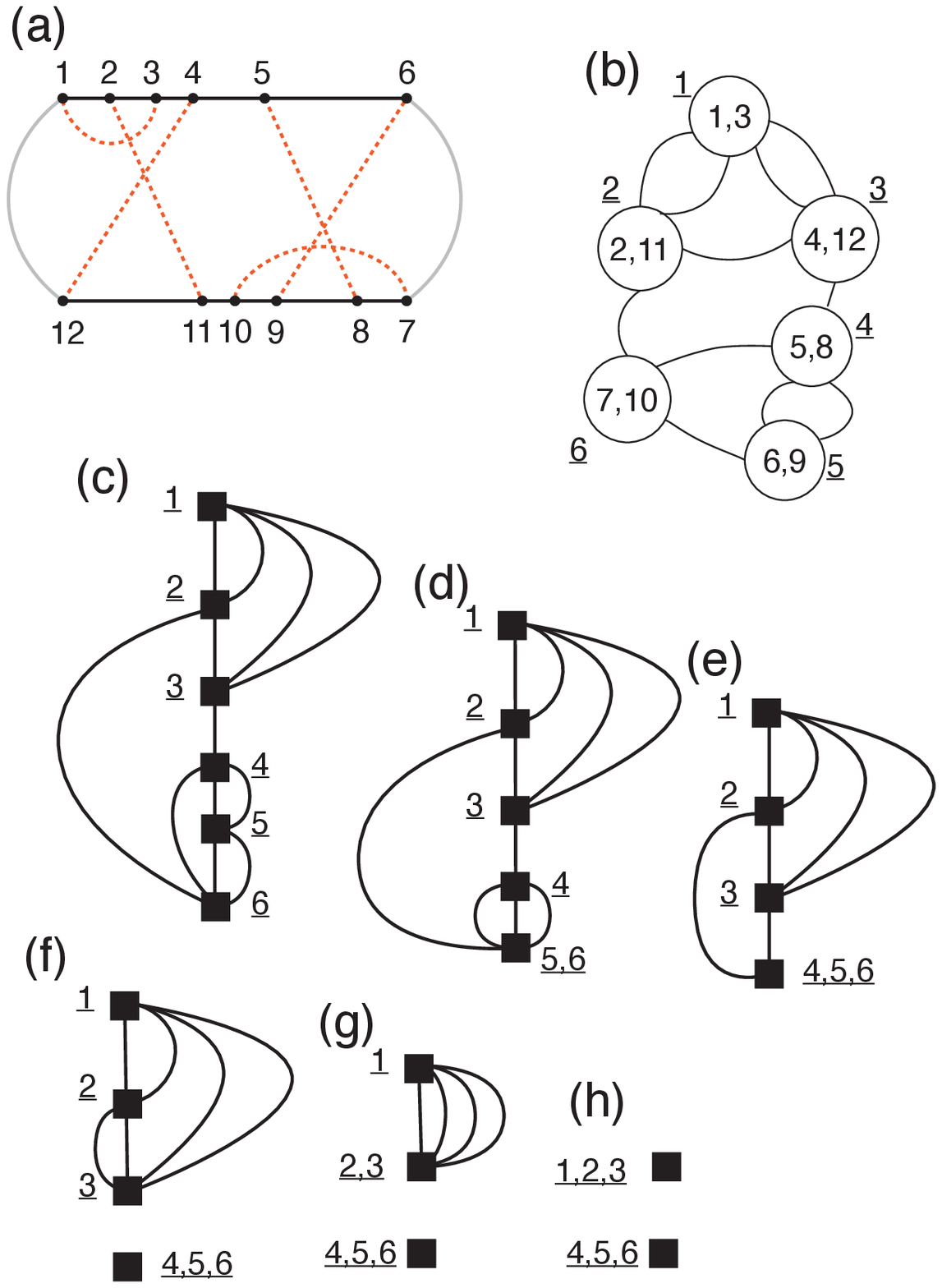}
	\caption{
		Algorithm for checking the 2PI property of the $Q$ vertex. 
		First, the original $Q$ vertex diagram shown in panel (a) is transformed into the graph structure of panel (b) 
		by combining the two vertices connected by red dashed lines into supervertices (circles). Here, the black lines represent PP propagators.  
		Note that in panel (a) we introduced auxiliary propagator lines (gray) connecting the vertices $1$ and $12$, as well as $6$ and $7$ in order to make the lowest-order diagram 2PI without affecting the 2PI property of higher-order diagrams.
		Panel (c) shows the tree representation of diagram (b) produced by the depth-first-search algorithm. 
		Panels (d-h) illustrate the absorption and ejection operations which allow to identify the separate three-edge-connected components.
	}
	\label{fig:2pi}
\end{figure}

A general Feynman diagram can be represented by a graph structure consisting of edges and vertices.
If the original $v$ vertices connected by $V$ lines are encapsulated into supervertices, the $Q$ diagrams which are 2PI in the $\mathcal{G}$ channel correspond to so-called three-edge-connected graphs. 
Figure~\ref{fig:2pi}(a) shows an example of a $Q$ diagram with 12 $v$ vertices, and Fig.~\ref{fig:2pi}(b) the corresponding graph with supervertices. 
To check for three-edge-connectivity,  
there exist several algorithms whose computational complexity scales as $\mathcal{O}(|E|)$, where $|E|$ is the number of edges in the graph.
We use the algorithm introduced in Ref.~\onlinecite{Norouzi2014}, which is based on a depth-first search of the graph.

In this algorithm, a given two-edge-connected graph (1PI Feynman diagram in the $\mathcal{G}$ channel) is successively transformed into a set of three-edge-connected components via absorption and ejection operations.
Whenever the depth-first search backtracks, the cardinality (the number of neighboring vertices) of the descendant 
(the last supervertex on the line) 
is investigated.
If it is $2$, the descendant is ejected as an isolated three-edge-connected component. Otherwise the graph is transformed using the absorption operation and the search proceeds.
Since we are only interested in the three-edge connectivity itself, we stop the checking routine after the first ejection operation.

Figure~\ref{fig:2pi}(c-h) illustrate how the graph in Fig.~\ref{fig:2pi}(b) is transformed.
Panel (c) shows the structure of the graph after the depth-first search is finished and the tree (vertical line) has been formed.
Next, we backtrack the tree while investigating the cardinality.
At the node $\underline{5}$, the cardinality of $\underline{6}$ is investigated. It is 4 and hence larger than 2. Thus, 
the node $\underline{6}$ is absorbed into $\underline{5}$, forming the supernode $\underline{5,6}$, as illustrated in panel (d). 
The absorption continues in the next step, since also the cardinality of node $\underline{5,6}$ is larger than 2, and we end up with the graph shown in panel (e).  
At the next step, we check the cardinality of the supernode $\underline{4,5,6}$~, which is 2. 
Now the ejection operation is applied and results in the disconnected graph shown in panel (f).
Such absorption and ejection operations are successively applied until the end (panels (g) and (h)). 
In the actual Monte Carlo sampling, we would stop the 2PI irreducibility checking at Fig.~\ref{fig:2pi}(f), since the existence of a disconnected part implies that the original graph was not 2PI.

\section{Dubiner basis representation}\label{sec:Dubiner}

Storing the $Q$ vertex is a challenging task because of the large number of data points generated by three different time indices and four different PP indices.
With a uniform time grid with $N_\tau$ time points and the size of the impurity Hilbert space $N_H$, the total number of data points required to store the $Q$ vertex scales as $N_H^{4}\times N_\tau(N_\tau+1)(N_\tau+2)/6$. As we increase $N_\tau$, the required memory quickly exceeds practical limits, for example, it is already $\sim 2.8$~GB for $N_H=4$ and $N_\tau=200$. In this section, we introduce a polynomial basis that can significantly reduce the memory cost for storing the $Q$ vertex. 

\begin{figure}[t]
	\centering
	\includegraphics[width=0.49\textwidth]{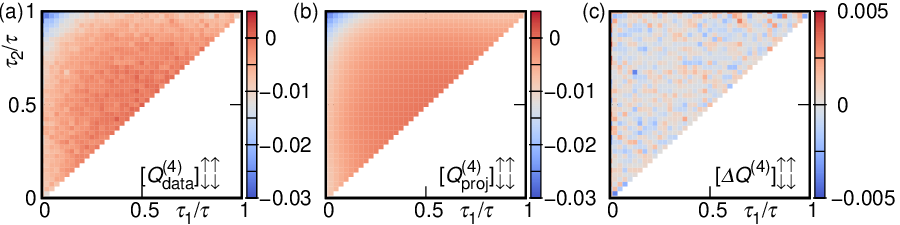}
	\caption{
		(a) The Monte Carlo sampled $[Q^{\uparrow\uparrow}_{\downarrow\downarrow}]^{(4)}(\tau=\beta/2,\tau_2,\tau_1)$ of the single-bath Anderson impurity model (Sec.~\ref{sec:benchmarkSingleBath}) for $T=0.1, U=1, \mu=0.5, V=0.5, \varepsilon=0$ and a time grid with $N_\tau=80$.
		(b) The reproduced $[Q^{(4)}]^{\uparrow\uparrow}_{\downarrow\downarrow}(\tau=\beta/2,\tau_2,\tau_1)$ after projection onto the Dubiner basis and truncation of the coefficients at basis order $5$.
		(c) The difference between the sampled and reproduced $[Q^{(4)}]^{\uparrow\uparrow}_{\downarrow\downarrow}(\tau=\beta/2,\tau_2,\tau_1)$.
}
	\label{fig:dubinerBenchmark}
\end{figure}

As one can see in Fig.~\ref{fig:dubinerBenchmark}(a), for fixed $\tau$ ($=\beta/2$ in the figure), $Q(\tau,\tau_2,\tau_1)$ is a smooth function of $(\tau_2,\tau_1)$.
Furthermore, because of the time ordering of $\tau_2$ and $\tau_1$ the domain of the function has a triangular shape. 
For a more efficient representation of this function, we thus use a polynomial basis adapted to this triangular time domain, the so-called Dubiner basis.\cite{Dubiner1991,Sherwin1995}

\begin{figure}[]
	\centering
	\includegraphics[width=0.49\textwidth]{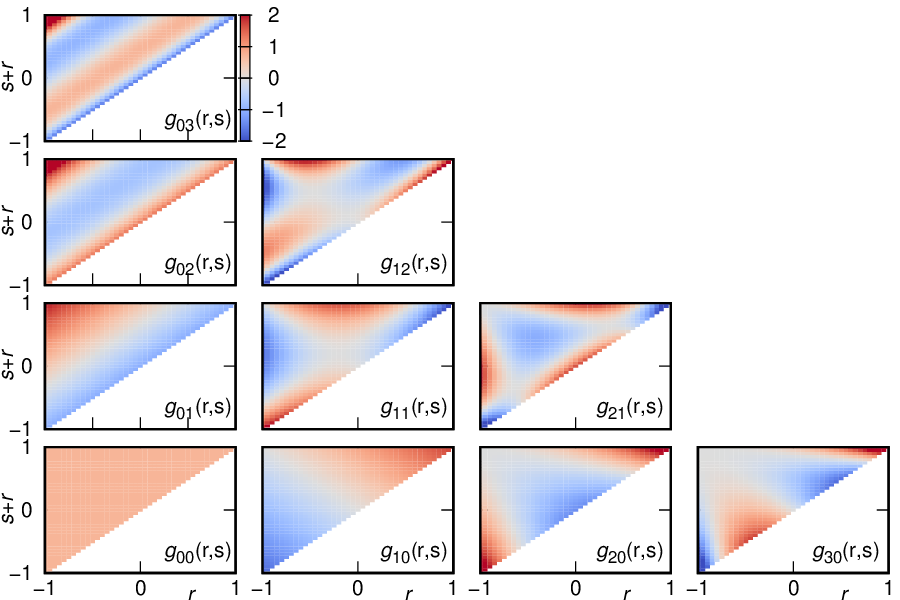}
	\caption{Dubiner basis $g_{lm}$ for $0 \le l + m \le 3$, which allows to represent polynomials up to third order.}
	\label{fig:Dubiner}
\end{figure}

The Dubiner basis is composed of a product of two Jacobi polynomials,
\begin{align}
	g_{lm}(r,s) =& \sqrt{\frac{(2l+1)(m+l+1)}{2^{2l+1}}}P^{0,0}_l\left(2\frac{1+r}{1-s}-1\right)\nonumber\\
	&\times \left(1-s\right)^lP^{2l+1,0}_m(s)~,
	\label{eqn:DubinerBasis}
\end{align}
where $-1\leq r,s;r+s\leq 0$.  Note that $g_{lm}$ constitutes a polynomial function whose maximum order is $l+m$.
Using the orthogonality relation 
\begin{equation}
	\int_{}^{}g_{lm}(r,s)g_{pq}(r,s)drds = \delta_{lp}\delta_{mq}~,
	\label{eqn:DubinerOrthonormality}
\end{equation}
one can directly accumulate the coefficients of the Dubiner basis during the MC sampling.
Figure~\ref{fig:Dubiner} shows the low-order basis functions up to polynomial order 3.

\begin{figure}[]
	\centering
	\includegraphics[width=0.45\textwidth]{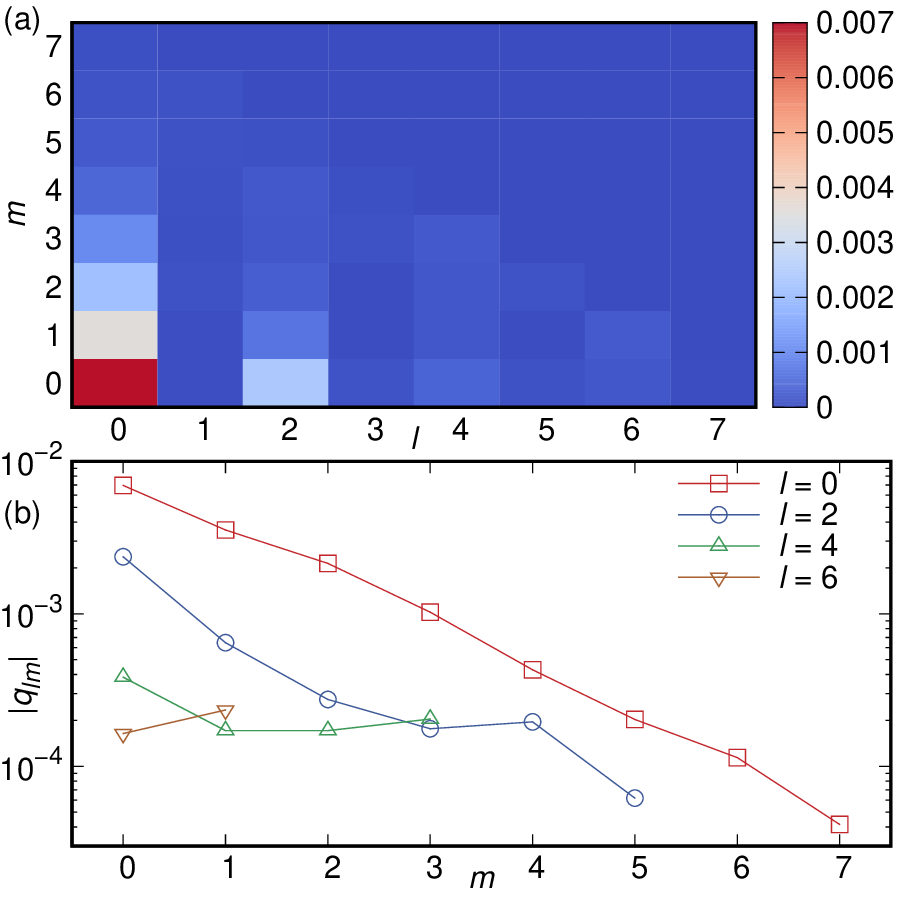}
	\caption{
		Exponential decay of the absolute value of the coefficients $q_{lm}$ of the Dubiner basis representation of the $Q$ vertex: 
		$[Q^{(4)}]^{\uparrow\uparrow}_{\downarrow\downarrow}(\tau=\beta/2;\tau_2,\tau_1) = \sum^{}_{lm}q_{lm}g_{lm}(2\tau_1/\tau-1,2(\tau_2-\tau_1)/\tau-1)$. The model and parameters are the same as in Fig.~\ref{fig:dubinerBenchmark}.
	}
	\label{fig:gcoeffBenchmark}
\end{figure}

For the purpose of illustration, we project the $[Q^{(4)}]^{\uparrow\uparrow}_{\downarrow\downarrow}(\tau=\beta/2;\tau_2,\tau_1)$ data of an Anderson impurity model obtained by the Monte Carlo sampling (Fig.~\ref{fig:dubinerBenchmark}(a)) onto the Dubiner basis:
\begin{equation}
	\!\!\left[Q^{(4)}\right]^{\uparrow\uparrow}_{\downarrow\downarrow}\!\!\Big(\tau=\frac{\beta}{2};\tau_2,\tau_1\Big) = \sum^{}_{lm}q_{lm}g_{lm}\Big(2\frac{\tau_1}{\tau}-1,2\frac{\tau_2-\tau_1}{\tau}-1\Big).
	\label{eqn:DubinerProjection}
\end{equation}
The coefficients $q_{lm}$ of the basis functions $g_{lm}$, shown in Fig.~\ref{fig:gcoeffBenchmark}, decay exponentially as a function polynomial order, which means that only a small number of coefficients is needed to represent the function. (The small upturns, for example in $q_{6m}$, are due to finite imaginary-time grids. See Appendix \ref{app:NtDepDubinerCoeff}.)
After truncating the coefficients beyond a cutoff value determined by the desired accuracy, we can reproduce the original data with the Monte Carlo noise filtered out.
Figure~\ref{fig:dubinerBenchmark}(b) presents the reproduced $[Q^{(4)}]^{\uparrow\uparrow}_{\downarrow\downarrow}(\tau=\beta/2;\tau_2,\tau_1)$ and Fig.~\ref{fig:dubinerBenchmark}(c) shows the difference between the sampled data and the reproduced data for a truncation at polynomial order $5$ (with $q_{lm}$ set to zero for $l+m>5$). With this truncation the error produced by the projection onto the Dubiner basis is of the order of $\sim 10^{-3}$, and thus smaller than the Monte Carlo noise. 

The Dubiner basis introduced above for the triangular time domain can be generalized to a tetrahedral domain.
In this case, the basis function is composed of the product of three Jacobi polynomials,
\begin{align}
	g_{lmn}(r,s,t)&=P^{0,0}_{l}\left(2\frac{1+r}{-s-t}-1\right)\left(\frac{-s-t}{2}\right)^l\nonumber\\
	&\times P^{2l+1,0}_{m}\left(2\frac{1+s}{1-t}-1\right)\left(\frac{1-t}{2}\right)^{m}\nonumber\\
	&\times P^{2l+2m+2,0}_{n}(t)~.
	\label{eqn:Dubiner3d}
\end{align}
Among several possibilities, one choice of $r,s,$ and $t$ is $r=2\tau_1/\beta-1, s=2(\tau_2-\tau_1)/\beta-1,$ and $t=2(\tau-\tau_2)/\beta-1$, respectively.

\section{Results}\label{sec:results}
\subsection{Single-orbital Anderson impurity model}\label{sec:benchmarkSingleBath}
\begin{figure}[t]
	\centering
	\includegraphics[width=0.45\textwidth]{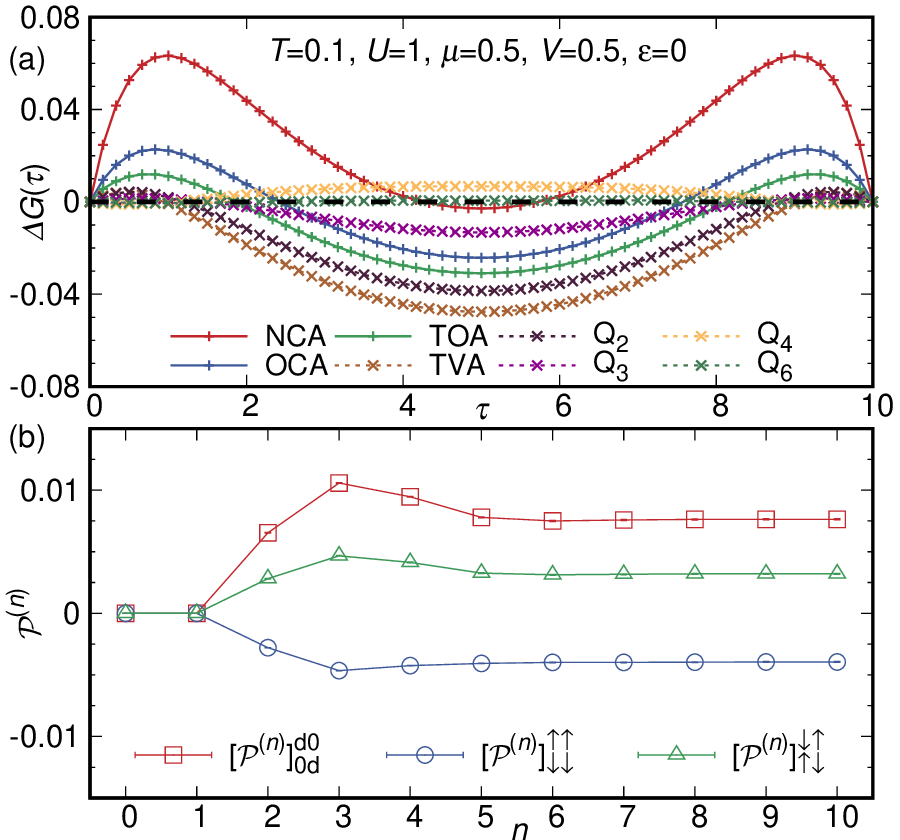}
	\caption{
		(a) Difference between the impurity Green's function from the indicated approximate scheme and the exact solution for the AIM with a single bath site. The parameters are $T=0.1$, $U=1$, $\mu=0.5$, $V=0.5$, and $\varepsilon=0$. The corresponding hybridization function is constant as a function of imaginary time.
		(b) The partial sum $\mathcal{P}^{(n)}$ of the $Q$ vertex series as a function of the maximum diagram order.  
		In these calculations, the self-consistent PP propagator of the $Q_{10}$ approximation is inserted into the PP vertex solver.
	}
	\label{fig:singleBathBenchmark}
\end{figure}

In this section, we benchmark the results of the PP vertex solver against exactly solvable models and investigate its convergence properties.
First, we study the single-bath AIM [Eq.~(\ref{eqn:Himp})], for which $F^\dagger_{\alpha=a}=d^{\dagger}_a$. It has a single bath degree of freedom with index $k=1$, while the flavor index $a$ represents the spin degrees of freedom $\uparrow$ and $\downarrow$.
The corresponding model parameters are $E_{ab} = -\mu\delta_{ab}$, $U_{ab,cd}=(U/2)\delta_{ad}\delta_{bc}$, 
 $V^{k=1}_{ab}=V\delta_{ab}$ and $\varepsilon_{k=1,ab}=\varepsilon\delta_{ab}$.

Figure~\ref{fig:singleBathBenchmark} illustrates the systematic convergence of the PP vertex solver as a function of the $Q$ vertex diagram order.
As a relevant observable, we consider the impurity Green's function $G(\tau)=G_{\uparrow\uparrow}(\tau)=G_{\downarrow\downarrow}(\tau)$.
Figure~\ref{fig:singleBathBenchmark}(a) presents the difference between the impurity Green's functions from several approximation schemes and the exact result, $\Delta G(\tau)=G_{\text{approx}}(\tau) - G_{\text{exact}}(\tau)$. The 
perturbative 
approximations (NCA, OCA, and TOA) exhibit sizable deviations from the exact Green's function with positive or negative signs, depending on $\tau$. While the self-consistent TVA is not obviously better, as we increase the maximum diagram order of the $Q$ vertex from 2 to 6, the Green's function clearly converges to the exact result.

When we achieve convergence in the set of self-consistent solutions, the 
partial sums of the terms contributing to the 
$Q$ vertex for a given PP propagator and PP interaction also converge as a function of diagram order. 
This is a stringent criterion for a convergent bold-diagrammatic series.\cite{Rossi2016} 
Figure~\ref{fig:singleBathBenchmark}(b) shows the partial sum of the series coefficients for the integrated $Q$ vertex for a given PP propagator: 
\begin{equation}
	\mathcal{P}^{(n)} = \sum_{k=0}^{n} \int^{\beta}_{0}d\tau\int_{0}^{\tau}d\tau_2\int_{0}^{\tau_2}d\tau_1Q^{(k)}(\tau;\tau_2,\tau_1)~.
	\label{eqn:intQpsum}
\end{equation}
Here, out of multiple channels for the external vertices, we only show those which are nonzero even for the lowest (second) order. 
The figure demonstrates an almost simultaneous convergence of each component around diagram order 6.

\begin{figure}[t]
	\centering
	\includegraphics[width=0.45\textwidth]{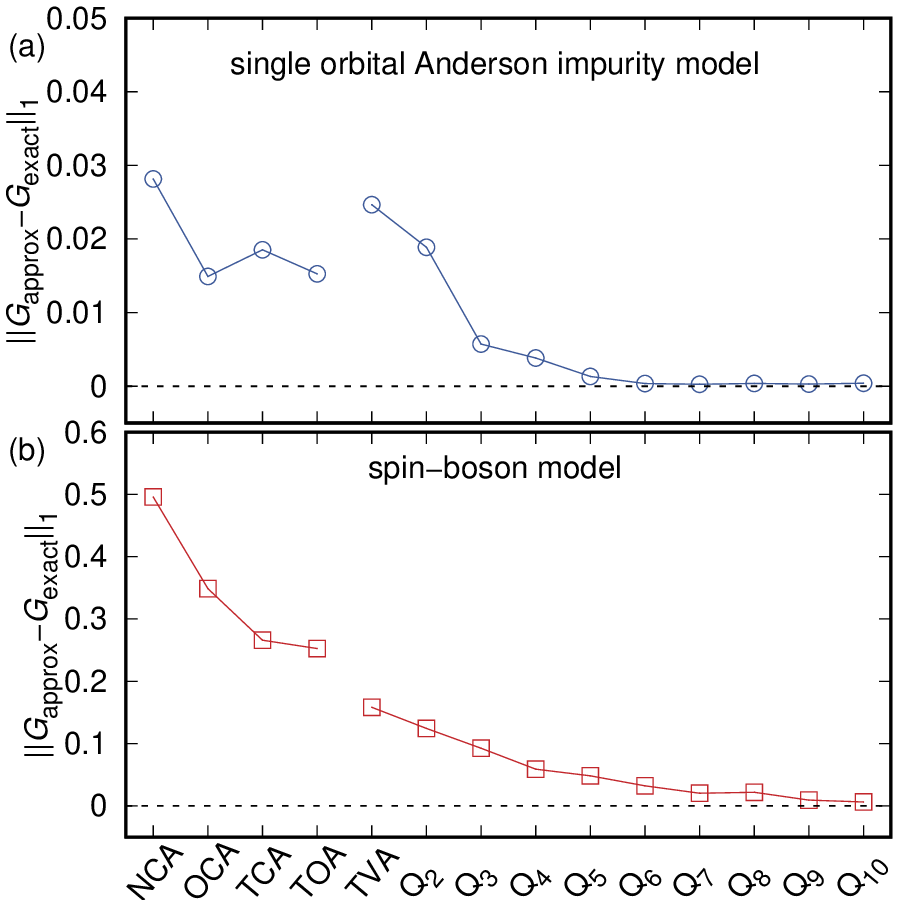}
	\caption{
		Average difference between the Green's function of the approximate scheme and the exact solution.
		(a) Single-bath AIM with $T=0.1, U=1, \mu=0.5, V=0.5$, and $\varepsilon=0$.
		(b) Single-boson mode ($k=1$) spin-boson model with $T=0.1, \mu=0.5, V=1.4$, and $\varepsilon=1.5$. The spin-boson model is defined in terms of Eq.~(\ref{eqn:Himp}) as follows: 
		$F_1=\sum_{ab}d^{\dagger}_{a}\sigma^x_{ab}d^{}_{b}$, 
		$E_{ab} = \mu\sigma^z_{ab}, V^{k=1}_{11}=V$, and $\varepsilon_{1,11} = \varepsilon$.
	}
	\label{fig:convergenceFermionBoson}
\end{figure}

An interesting question is how the particle statistics of the bath degrees of freedom and the complexity of the local impurity problem affect the convergence of the PP vertex solver.
As shown in Fig.\ref{fig:convergenceFermionBoson}(b),\cite{Aaram2021} in the case of the spin-boson model with bosonic bath degrees of freedom and a two-dimensional local Hilbert space,
the accuracy of the PP vertex solver monotonically improves as we sum up more diagrams. 
In particular, the triangular vertex self-consistency [Fig.~\ref{fig:scEqn}(c)], i.e., the step from TCA to TVA, considerably improves the accuracy of the scheme.

\begin{figure}[t]
	\centering
	\includegraphics[width=0.45\textwidth]{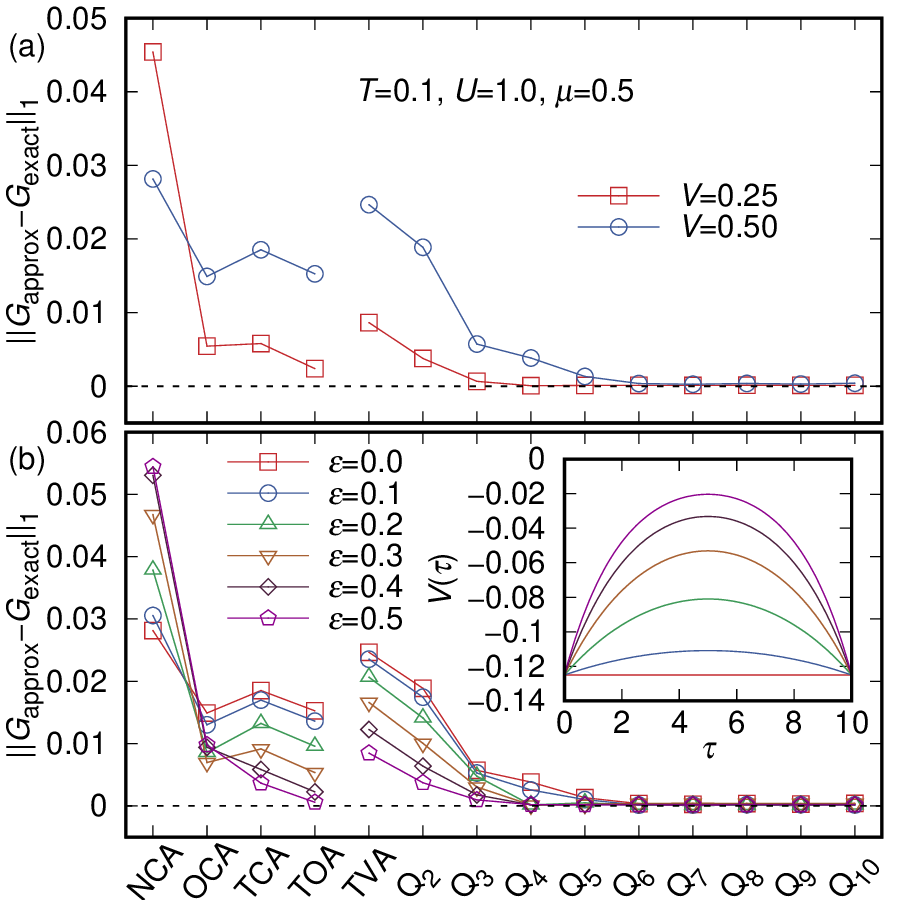}
	\caption{
		Average difference between the Green's function of the indicated approximate scheme and the exact solution for the AIM (a) with a single bath and two different $V$ values, and (b) with two baths at energy $\pm \varepsilon$. 
		The other parameters are $T=0.1$, $U=1$, and $\mu=0.5$.
		The inset in panel (b) shows the PP retarded interaction $V(\tau)$ for the different $\varepsilon$ values.
	}
	\label{fig:convergenceFermion}
\end{figure}

In the fermionic AIM, however, there is no simple monotonic convergence. This is already clear by comparing the results for OCA and TCA. Furthermore, 
for all parameter sets investigated, the TVA is less accurate than the TCA.  The main difference between the two schemes is the vertex self-consistency.
Figure~\ref{fig:convergenceFermion}(a) shows the average value of $|\Delta G(\tau)|$ for various approximate schemes.
Despite the eventual convergence of the self-consistent vertex scheme, the TVA and low-order $Q$ vertex schemes suffer from slow convergence, even compared to the bare vertex schemes such as OCA and TOA.

The initially slow convergence of the self-consistent vertex scheme becomes even more evident when we introduce two fermionic bath degrees of freedom, $\varepsilon_{k=1,ab}=+\varepsilon\delta_{ab}$ and $\varepsilon_{k=2,ab}=-\varepsilon\delta_{ab}$, which allows to 
produce a more realistic shape of the hybridization function, as shown in the inset of Fig.~\ref{fig:convergenceFermion}(b). 
In the main panel, we observe that the bare vertex schemes approach the exact results more rapidly than the vertex self-consistent schemes as we increase the $\varepsilon$ values.
In particular for weakly hybridized systems (rapidly decaying $V(\tau)$), the bare vertex scheme converges fast and seems to out-perform the vertex self-consistent schemes, which sum up many more diagrams, including rainbow-type vertex corrections.  As in the case of weak-coupling diagrammatic schemes,\cite{Gukelberger2015} the strategy of summing up certain sub-classes of diagrams to infinite order is not always optimal for the AIM.    

The obvious difference between the bosonic and fermionic models could be an indication that the convergence properties are influenced by the sign structure (periodic/anti-periodic) of the hybridization function; while the hybridization function of the bosonic bath is $\beta$-periodic and has a definite sign independent of the time argument, the hybridization function of the fermionic bath is $\beta$-antiperiodic, and thus its sign depends on the time arguments.
On the other hand, compared to the spin-boson model, the size of the local Hilbert space is twice as large in the AIM and the physics is more complex.  To really judge whether the sign structure of the hybridization function has an important effect on the convergence, one would have to compare the spin-boson model to the spinless Anderson impurity model. 

\begin{figure}[t]
	\centering
	\includegraphics[width=0.45\textwidth]{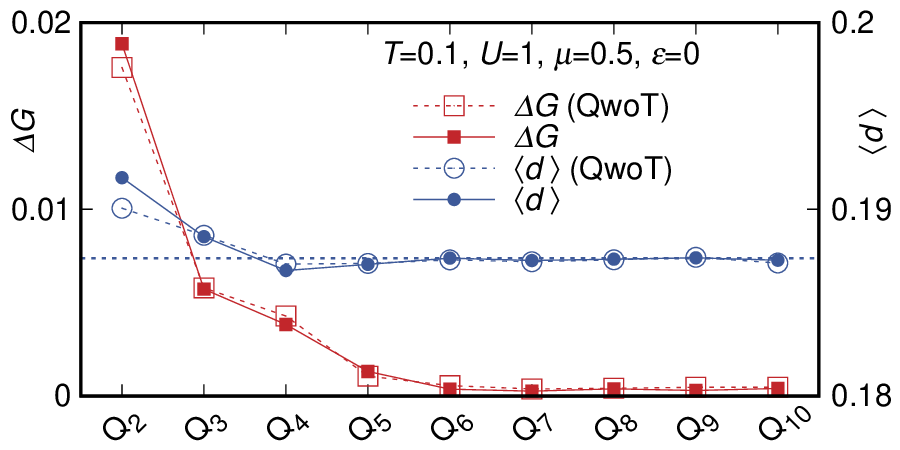}
	\caption{
		Average difference between the imaginary-time Green's function of the indicated approximation scheme and the exact value (red symbols, left axis)  and double occupancy obtained from the pseudo-particle Green's function (blue symbols, right axis) for $T=0.1, U=1, \mu=0.5, V=0.5,$ and $\varepsilon=0$.
		Solid [open] symbols are obtained using the self-consistent vertex scheme of Fig.~\ref{fig:scEqn}(c) [Fig.~\ref{fig:vertexsc_woT}(a)].
		The horizontal dashed line indicates the exact double occupancy.
	}
	\label{fig:QwoT}
\end{figure}

One might suspect that the possible overdressing of the triangular vertex in the vertex self-consistency equation is the origin of the slow convergence observed for the AIM.
Note that the $T_2, T_3, T_4, T_6, T_7$, and $T_8$ terms in Fig.~\ref{fig:scEqn}(c) include a \textit{dressed} triangular vertex in the 2PI part, which is typically absorbed into the four-point vertex.
In order to identify the effect of the dressed triangular vertex in the 2PI part, we compare the convergence of two different self-consistent schemes: with and without (QwoT) the dressed triangular vertex, as detailed in Appendix~\ref{app:QwoT}.
It turns out that the scheme without the dressed triangular vertex in the 2PI part produces a qualitatively very similar convergence behavior to the one with the fully dressed triangular vertex; see Fig.~\ref{fig:QwoT}. This implies a different origin of the slow convergence, and in particular of the poor performance of the TVA and $Q_2$ approximations. 
Although the double-occupancy result of the QwoT scheme converges slightly faster, the difference is not significant.

One relevant observation is that for $n\ge 3$ the $Q_n$ approximation actually does converge rapidly. 
This suggests that the poor performance of the TVA and $Q_2$ approximation originates from an oversimplified ladder structure. 
TVA sums up conventional ladder-type diagrams, while $Q_2$ additionally takes into account single crossings between rungs. 
In a systematic diagMC study within the conventional weak-coupling diagrammatic framework, it was found that the resummation of such simple ladder diagrams to infinite order does not produce accurate results.\cite{Gukelberger2015} 
Ladder-type diagrams with complicated topologies (multiple crossings) are as important as those with simple topologies. 
A related problem appears when one estimates the Kondo coupling within the NCA, 
since crossing and noncrossing diagrams of the same order contribute equally to the PP self-energy.~\cite{Pruschke1989}
Our results for the TVA and $Q_2$ schemes suggest that these issues also affect the present strong-coupling diagrammatic framework. 
We need the $Q_n$ diagrams with $n\ge3$ to capture the relevant cancellations between different ladder topologies.  

\subsection{DMFT results for the Hubbard model}

The PP vertex solver can be used as an impurity solver in the DMFT self-consistent equations.
We thus also benchmark the PP vertex solver with paramagnetic DMFT solutions.
As a testbed, we consider the single-orbital Hubbard model on the infinite-dimensional Bethe lattice with semicircular density of states, $\rho(\varepsilon) = (2/\pi D)\sqrt{1-(\varepsilon/D)^2}$, where $D$ is the half-bandwidth.
Within DMFT, this Hubbard model is mapped onto the AIM subject to the self-consistency condition $V_{\text{DMFT}}(\tau)=(D^2/4)G(\tau)$.\cite{Georges1996}
To test the accuracy of the PP vertex solver, we first determine the hybridization function of the exact DMFT solution, $V_{\text{DMFT}}(\tau)$, using CT-HYB, and then use this and the corresponding exact $G(\tau)$ in the benchmark calculations. 

\begin{figure}[]
	\centering
	\includegraphics[width=0.45\textwidth]{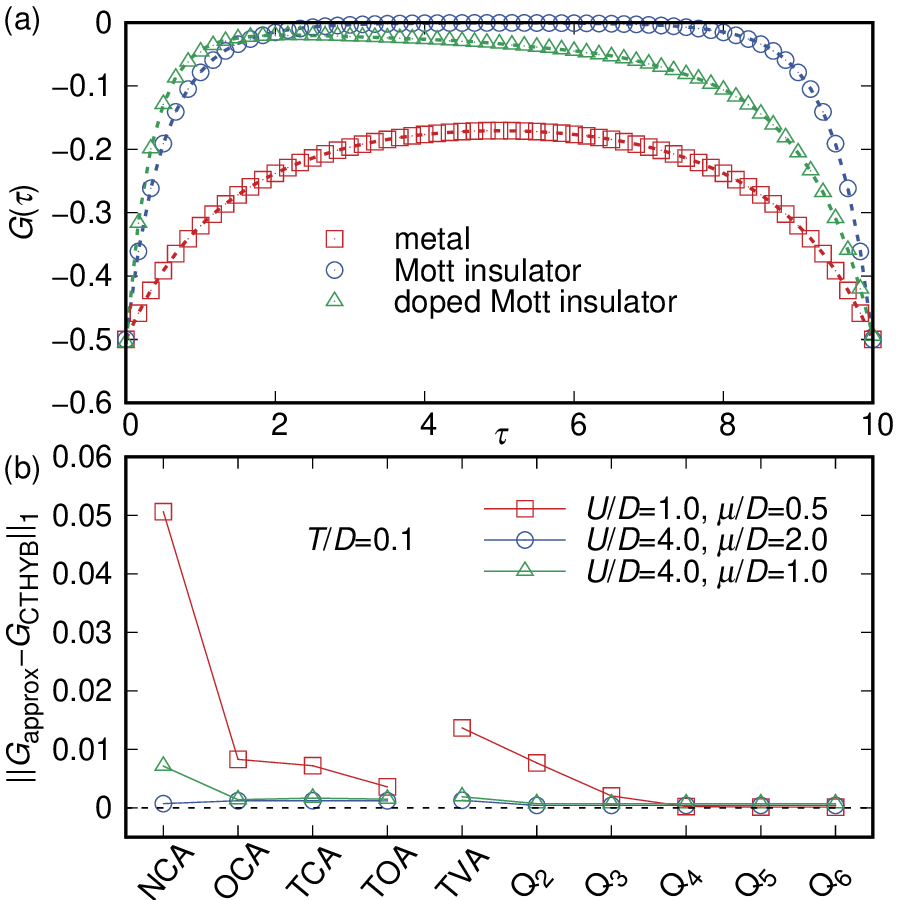}
	\caption{
		DMFT benchmarks for three different regimes of the single-band Hubbard model: metal, Mott insulator, and doped Mott insulator.
		The temperature is fixed to $T=0.1D$, where $D$ is the half-bandwidth of the semicircular density of states.
		(a) Open symbols present the results obtained by the PP vertex solvers and the dashed lines the CT-HYB results.
		(b) Average deviation of the impurity Green's function from the CT-HYB result, for the indicated approximations. 
	}
	\label{fig:dmftBenchmark}
\end{figure}

Figure~\ref{fig:dmftBenchmark}(a) shows the obtained paramagnetic impurity Green's functions for three different parameter regimes: metal, Mott insulator, and doped Mott insulator.
For all those solutions, we achieve convergence as a function of $Q$ vertex diagram order.
As one can expect {\it a priori}, the convergence 
is faster 
for a more localized system.
While in the Mott insulator, the low-order approximations such as OCA, TCA, and TOA already provide very good approximations of the CT-HYB solution, high-order $Q$ vertex contributions are essential for achieving the same accuracy in the metallic phase.
The doped Mott insulator shows a convergence that is faster than in the metal but slower than in the case of the Mott insulator.

Overall, the convergence behavior as a function of the $Q$ vertex diagram order is similar to the single- or two-bath AIM discussed in Sec.~\ref{sec:benchmarkSingleBath}.
The bare vertex schemes, such as TCA and TOA, behave better than the corresponding schemes with vertex self-consistency, TVA and $Q_2$, in all three phases; see Fig.~\ref{fig:dmftBenchmark}(b). 
For $Q_n$ with $n\ge 3$, the self-consistent vertex scheme also becomes accurate. 

\begin{figure}[]
	\centering
	\includegraphics[width=0.45\textwidth]{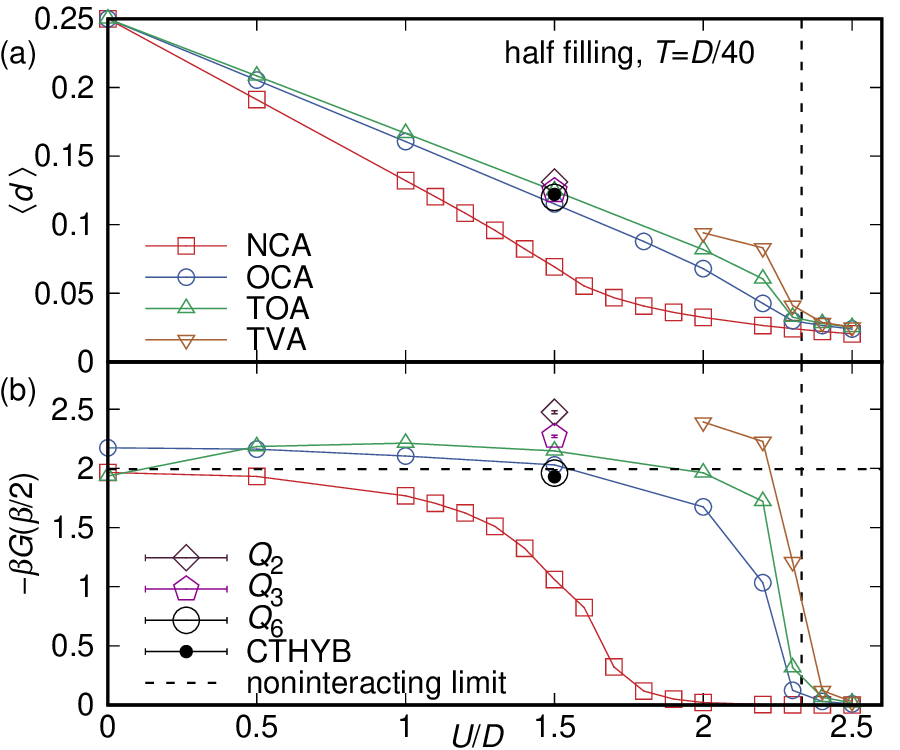}
	\caption{
		(a) Double occupancy $\langle d\rangle$ and (b) approximate spectral function at the Fermi level $\bar{A}(\omega=0)=-\beta G(\beta/2)$ of the half-filled single-band Hubbard model.
		The temperature is fixed as $D/40$, where $D$ is the half-bandwidth of the semi-circular density of states.
		Results which include high-order vertex corrections are presented for $U=1.5D$.
		The vertical dashed lines correspond to the interaction strength $U_c$ of the Mott transition 
		point estimated by CT-HYB, and the horizontal line shows the noninteracting low-energy spectral function.
	}
	\label{fig:MottTransition}
\end{figure}
Finally, we consider the Mott transition of the single-band Hubbard model and compare the results from the various approximate schemes.
Figure~\ref{fig:MottTransition} presents the double occupancy and an estimate for the spectral function at the Fermi level, 
\begin{align}
	\bar{A}(\omega=0)&\equiv -\beta G(\beta/2)\nonumber\\
	&=\beta\int_{-\infty}^{\infty}d\omega~\frac{A(\omega)}{2\cosh\left(\beta\omega/2\right)}~,
\end{align}
as a function of the on-site interaction $U$.
The first noticeable observation is the strong underestimation of the critical interaction strength $U_c$ and the finite-temperature end point of the Mott transition line $T_c$ in the NCA.
The observed $U_c^{\textrm{NCA}}\sim 1.65$ is around 30\% smaller than the known exact value $\sim 2.33$ at the end point $T^{\mathrm{NCA}}_c\sim 0.0225$, which  is significantly lower than the exact $T_c\sim 0.027$.\cite{Kim2014}
This overestimation 
of the correlation effects can be attributed to the missing exchange processes between the impurity and bath, which has been a main motivation for the development of higher-order corrections like OCA and the TOA.
While the OCA and the TOA estimate the critical interaction strength and temperature with rather high precision, local quantities such as the double occupancy and low-energy observables such as $\bar{A}(\omega=0)$ are still far from converged. 
As one can see in Fig.~\ref{fig:MottTransition}(b), for example, the OCA and the TOA considerably overestimate the low-energy spectral function in the intermediate and low-correlation regime. 
Particularly, the TOA shows an unphysical increase of $\bar{A}(\omega=0)$ with increasing $U$ in the small-$U$ regime. 
This overestimation becomes even worse in the low-order vertex schemes, e.g., $Q_2$ at $U=1.5D$, while the high-order vertex corrections systematically improve the results toward the exact reference value obtained by the CT-HYB algorithm.

\begin{figure*}[ht]
	\centering
	\includegraphics[width=1.0\textwidth]{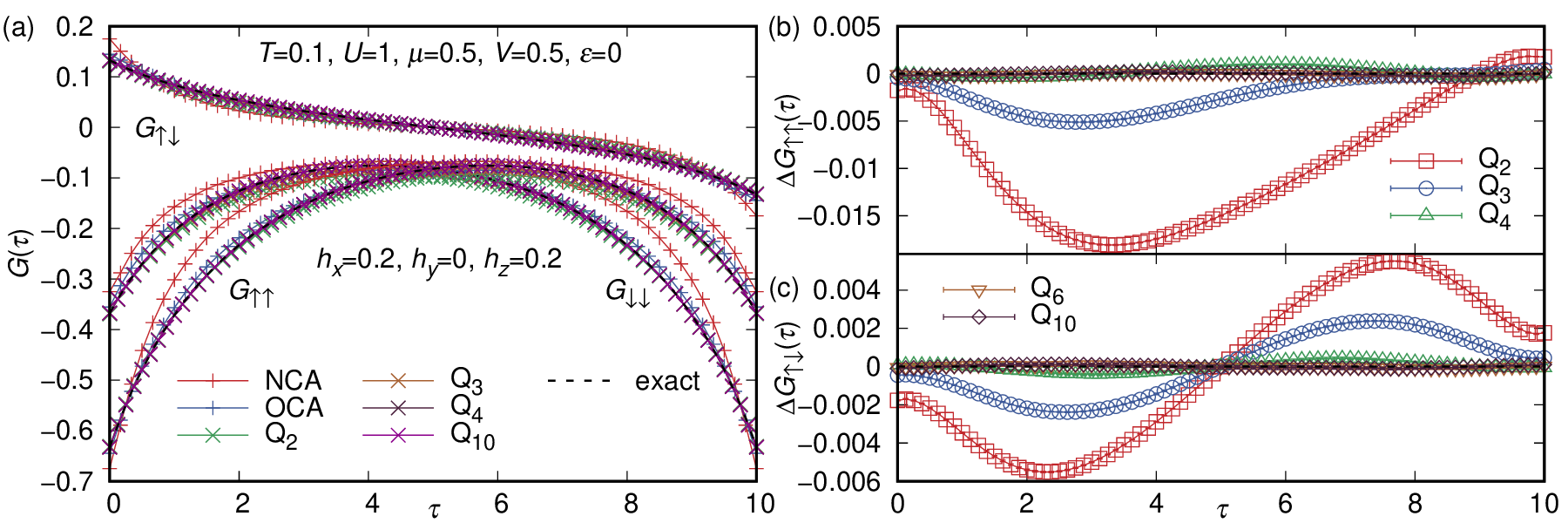}
	\caption{
		Benchmarks for the model with broken spin-rotational symmetry.
		(a) Diagonal ($G_{\uparrow\uparrow}$ and $G_{\downarrow\downarrow}$) and off-diagonal ($G_{\uparrow\downarrow}$ or $G_{\downarrow\uparrow}$) components of the impurity Green's function for $T=0.1$, $U=1$, $\mu=0.5$, $V=0.5$, and $\varepsilon=0$.
		We apply the magnetic field to the bath degrees of freedom along the $zx$ direction: $h_x=h_z=0.2$ and $h_y=0$.
		Panels (b) and (c) present the difference between the results of the various $Q$-vertex approximations and the exact solution for $G_{\uparrow\uparrow}$ and $G_{\uparrow\downarrow}$, respectively.
	}
	\label{fig:sbBenchmark}
\end{figure*}

\subsection{Off-diagonal hybridizations}

It is interesting to see how the PP vertex solver performs in the simulation of a model with a potential sign problem. 
For this benchmark we consider the 
AIM with an arbitrarily oriented magnetic field acting on the impurity spin, which can produce off-diagonal hybridization functions and a serious sign problem in the CT-HYB algorithm. 
To the AIM described in Sec.~\ref{sec:benchmarkSingleBath}, we add a Zeeman term $\mathbf{h}\cdot\bm{\sigma}$, so that $\varepsilon_{k=1,ab}=\varepsilon\delta_{ab}+(\mathbf{h}\cdot\bm{\sigma})_{ab}$. Such a term may appear in the DMFT impurity model of a lattice problem with external magnetic field, and it is known to generate a serious sign problem in CT-HYB\cite{Eidelstein2020} in a representation with non-zero off-diagonal hybridization functions $\Delta_{\uparrow\downarrow}$ (or $\Delta_{\downarrow\uparrow}$) and PP Green's functions $\mathcal{G}_{\uparrow\downarrow}$ (or $\mathcal{G}_{\downarrow\uparrow}$).
Although one could diagonalize the hybridization function via a simple basis transformation in this case, we use here a representation with off-diagonal components in order to investigate the convergence properties of the vertex solver in a situation which is challenging for other Monte Carlo solvers. 

Figure~\ref{fig:sbBenchmark} presents the benchmark results for the impurity Green's function in the presence of the magnetic field $h$ along the $zx$-direction.
In Fig.~\ref{fig:sbBenchmark}(a), we confirm that by including high-order $Q$ vertex contributions (up to order $10$) the PP vertex solver reproduces the exact results.
The diagram orders where convergence is achieved are approximately the same as in the model without magnetic field.
In panels (b) and (c), which plot the difference to the exact solution, one can see that the $Q_6$ approximation already reproduces the exact results within an error of order $10^{-4}$ for both the diagonal and off-diagonal components of the Green's function.
The fact that there is no indication of performance degradation shows that the vertex solver does not suffer from a conventional fermionic sign problem. Together with the fact that the vertex scheme works with connected diagrams, this suggests that it should be promising for tackling the sign (phase) problem which appears on the real-time axis, and which is very severe in CT-HYB.\cite{Werner2009} Hence, the vertex solver developed here could be potentially useful for applications to nonequilibrium systems.

\section{Conclusions}
In this paper, we described in detail the pseudo-particle vertex solver for quantum impurity models, which has recently been used in Ref.~\onlinecite{Aaram2021}  
to study a two-level system in a waveguide (spin-boson problem). We formulated the method for generic impurity models and tested it on the spin-boson model and various types of Anderson impurity models. 

In the pseudo-particle vertex solver, we sample the four-point vertex function using a diagrammatic Monte Carlo method and plug it into a self-consistent equation for the triangular vertex. Since the algorithm is based on a time-stepping procedure which successively extends the vertex $T(\tau,\tau')$ in a two-dimensional time domain, an appropriate nickname may be the slime mold algorithm.
The triangular vertex in turn defines the pseudo-particle self-energy and the pseudo-particle Green's function. With these updated Green's functions, a new four-point vertex is calculated and the procedure is repeated until a converged solution is found. 
With the converged triangular vertex and the pseudoparticle Green's functions, the slime mold algorithm gives direct access to two-time observables such as the impurity Green's function, in contrast to the inch-worm algorithm which requires an additional costly simulation for the vertex components.~\cite{Antipov2017}

In the specific implementation of the diagMC sampling, we adopt an efficient graph algorithm that filters out non-2PI diagrams with a computational effort that scales linearly with the diagram order. 
This algorithm can be used more broadly to construct diagrams composed of the dressed interaction or Green's function lines, e.g., in the bold-line diagrammatic Monte Carlo method.\cite{Prokofev2007,Prokofev2008}
We also showed that a polynomial basis for the tetrahedral time domain, the so-called Dubiner basis, can store the four-point vertex function with three imaginary-time arguments in a compact fashion.  
For nonequilibrium applications, it will be interesting to explore if the Dubiner basis can efficiently store the vertex functions and Green's functions of nonequilibrium systems measured on the real-time axis.

We benchmarked the pseudo-particle vertex solver by considering various exactly solvable models, and confirmed that it converges to the exact results as we increase the maximum diagram order of the four-point vertex.
We also confirmed that it can be used as an impurity solver for dynamical mean-field theory calculations.
Importantly, we demonstrated that the vertex solver handles impurity models with off-diagonal hybridizations without a loss of performance, even though this situation leads to a serious sign problem in standard continuous-time quantum Monte Carlo methods which are based on a partition function expansion.

For very large pseudoparticle interactions (larger than the interactions considered in this paper), we observed indications of a multivaluedness problem\cite{Kozik2015} in a pseudo-particle diagrammatic formalism.
Figuring out under which conditions such a multivaluedness problem is encountered will be an interesting topic for a future investigation. 

\acknowledgements

The calculations have been run on the Beo05 cluster at the University of Fribourg. A.K. and P.W. acknowledge support from ERC Consolidator Grant No. 724103 and M.E. from ERC Starting Grant No. 716648. J.L. is supported by SNSF Grant No. 200021-196966 and Marie Sklodowska Curie Grant Agreement No. 884104 (PSI-FELLOW-III-3i).

\begin{figure}[t]
	\centering
	\includegraphics[height=0.5\textwidth,angle=270]{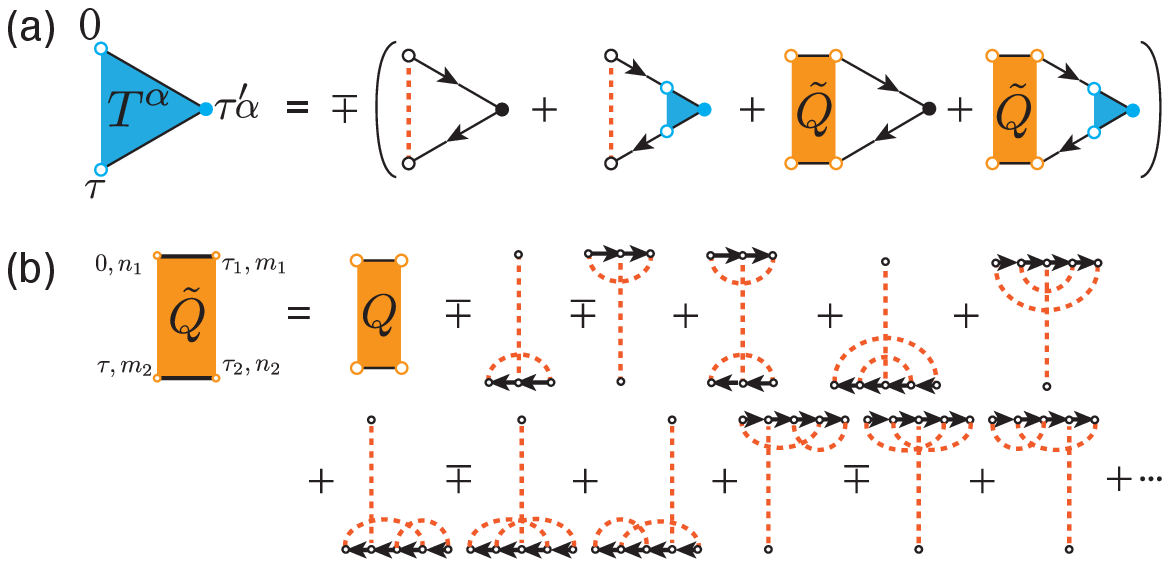}
	\caption{
		(a) Triangular vertex self-consistency equation of the QwoT scheme. 
		(b) Low-order $\tilde Q$ vertex diagrams up to the third order.
	}
	\label{fig:vertexsc_woT}
\end{figure}

\appendix
\section{QwoT scheme}\label{app:QwoT}

In this appendix, we describe an alternative vertex self-consistency equation, which does not include any dressing by a triangular vertex of the single vertical interaction line that connects the upper and lower backbone propagator. Instead, such vertex corrections are included explicitly (at each given order) in the modified four-point vertex $\tilde{Q}$, as illustrated in Fig.~\ref{fig:vertexsc_woT}.
In particular, Fig.~\ref{fig:vertexsc_woT}(b) presents all the additional four-point vertex diagrams up to the third order.
The diagMC solver has been modified to sample the $\tilde{Q}$ vertex directly. The 1PI condition for the interaction lines is released and we separately sample the $\tilde{Q}$ diagrams with a single vertex on the upper or lower segment in order to take into account the resulting delta function between the external time arguments. 

\section{Imaginary-time grid dependence of the Dubiner coefficients}\label{app:NtDepDubinerCoeff}
\begin{figure}[]
	\centering
	\includegraphics[width=0.45\textwidth]{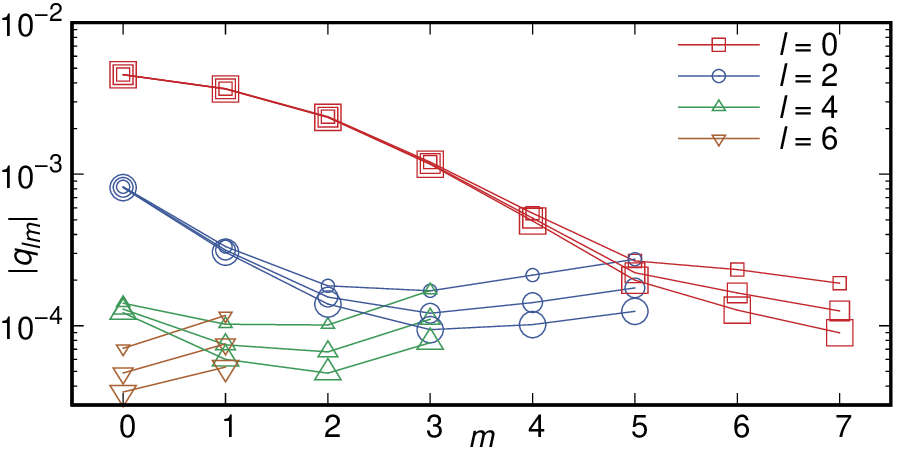}
	\caption{
		Imaginary-time grid dependence of the Dubiner coefficients for the exact $Q_2$ vertex.
		The three different symbol sizes represent the number of time-grid points: 80, 100, and 120 points, respectively.
		The model and parameters are the same as in Fig.~\ref{fig:dubinerBenchmark}.
	}
	\label{fig:gcoeffBenchmark_Q2}
\end{figure}

Although the coefficients $q_{lm}$ from the projection of the $Q$ vertex onto the Dubiner basis functions $g_{lm}$ exponentially decrease as a function of $m$ for large $m$ and fixed $l$, we observe some spurious effects related to the finite imaginary-time grid.
For example, the upturn in $|q_{l=6,m}|$ coefficients in Fig.~\ref{fig:gcoeffBenchmark}(b) can be attributed to this finite time-grid effect. 
In order to investigate the grid-size dependence without Monte Carlo error, we projected the exact $Q_2$ vertex\footnote{It is possible to exactly compute the $Q$ vertex only for low orders.} onto the Dubiner basis.
Figure~\ref{fig:gcoeffBenchmark_Q2} presents the Dubiner coefficients for different imaginary-time grids.
As we increase the number of grid points, the increasing $|q_{lm}|$ coefficients as a function of $m$ exhibit a strong grid-size dependence, while the converged coefficients shows a clear exponential decay.
We note, however, that these projection errors, for realistic time grids, are smaller than the stochastic Monte Carlo errors.

\bibliography{ref}

\end{document}